\documentclass[aps,a4paper,superscriptaddress,showpacs,preprintnumbers,amsmath,amssymb]{revtex4}

\usepackage{ulem}

\usepackage{psfrag} \usepackage{graphicx} \usepackage{dcolumn}
\usepackage{color} \usepackage{latexsym,amsfonts} \usepackage{bm}
\usepackage{amssymb}
\begin{document}

\title{Proton Recoil Energy and Angular Distribution of Neutron
  Radiative $\beta^-$--Decay}

\author{A. N. Ivanov}\email{ivanov@kph.tuwien.ac.at}
\affiliation{Atominstitut, Technische Universit\"at Wien, Stadionalle
  2, A-1020 Wien, Austria}
\author{R. H\"ollwieser}\affiliation{Atominstitut, Technische
  Universit\"at Wien, Stadionalle 2, A-1020 Wien, Austria}
\author{N. I. Troitskaya} \affiliation{State Polytechnic University of
  St. Petersburg, Polytechnicheskaya 29, 195251, Russian Federation}
\author{M. Wellenzohn}\affiliation{Atominstitut, Technische
  Universit\"at Wien, Stadionalle 2, A-1020 Wien, Austria}

\date{\today}

\begin{abstract}
We analyse the proton recoil energy and angular distribution of the
radiative $\beta^-$--decay of the neutron to leading order in the
large baryon mass expansion by taking into account the contributions
of the proton--photon correlations. We show that the account for the
proton--photon correlations does not contradict the description of the
radiative corrections to the lifetime of the neutron and the proton
recoil energy spectrum of the neutron $\beta^-$--decay in terms of the
functions $(\alpha/\pi)\,g_n(E_e)$ and $(\alpha/\pi)\,f_n(E_e)$, where
$E_e$ is the electron energy. In addition we find that the
contributions of the proton--photon correlations in the radiative
$\beta^-$--decay of the neutron to the proton recoil asymmetry $C$ are
of order $10^{-4}$. They make the contributions of the radiative
corrections to the proton recoil asymmetry $C$ symmetric with respect
to a change $A_0 \longleftrightarrow B_0$, where $A_0$ and $B_0$ are
the correlation coefficients of the neutron $\beta^-$--decay.
\end{abstract}

\maketitle

\section{Introduction}
\label{sec:introduction}

Recently contributions of order $10^{-4}$ of interactions beyond the
Standard model (SM) to the neutron $\beta^-$--decay have been
investigated in \cite{Ivanov2012}. For the analysis of such
contributions to the proton recoil energy spectrum $a(T_p)$, where
$T_p$ is a kinetic energy of the decay proton, and the proton recoil
asymmetry, defined by a correlation coefficient $C$, the calculation
of the proton recoil energy and angular distribution has been
performed by taking into account a complete set of corrections, caused
by the ``weak magnetism'' and the proton recoil, calculated to
next--to--leading order in the large baryon mass expansion, and the
radiative corrections of order $\alpha/\pi \sim 10^{-3}$, calculated
to leading order in the large baryon mass expansion
\cite{Ivanov2012}. For the cancellation of the infrared divergences,
caused by one--virtual photon exchanges, the standard procedure
\cite{Sirlin1967,Abers1969} has been used and the contribution of the
radiative $\beta^-$--decay of the neutron has been added
\cite{Ivanov2012}. The calculation of the contribution of the
radiative $\beta^-$--decay has been performed, first, by integrating
over of the proton 3--momentum and, second, by integrating over the
antineutrino energy. This has led to the radiative corrections,
defined by two functions $(\alpha/\pi)\,g_n(E_e)$ and
$(\alpha/\pi)\,f_n(E_e)$, which were calculated for the first time by
Sirlin \cite{Sirlin1967} and Shann \cite{Shann1971}, respectively.
However, as has been pointed out by Gl\"uck \cite{Gluck1997} (see also
\cite{Christian1978}), the calculation of the contribution of the
radiative $\beta^-$--decay of the neutron to the proton recoil energy
spectrum and the proton recoil asymmetry is much more complicated. For
the calculation of such a contribution one has to integrate, first,
over the antineutrino 3--momentum and then over other dynamical
variables. In spite of the calculation of the radiative corrections to
leading order in the large baryon mass expansion such an integration
leads to correlations between a recoil proton and an emitted photon,
i.e. the proton--photon correlations. The region of the integration
over the photon energy spectrum should be divided into two parts
\cite{Gluck1997}, corresponding to an emission of i) the soft-photons,
the contribution of which is responsible for a cancellation of the
infrared divergences of one--virtual photon exchanges in the neutron
$\beta^-$--decay, and of ii) the hard-photons, which define a part of
the observable radiative corrections to the neutron $\beta^-$--decay.
As has been pointed out by Gl\"uck \cite{Gluck1997}, the calculation
of the contributions of the hard-photons to $\beta$--decays may be
carried out only numerically. For the aim of the calculation of the
hard--photon corrections Gl\"uck has used the Monte Carlo simulation
method \cite{Gluck1997}.

In this paper we follow the paper by Gl\"uck \cite{Gluck1997} and
revise the contribution of the radiative $\beta^-$--decay of the
neutron, obtained in \cite{Ivanov2012}. We perform the integration
over the antineutrino 3--momentum and analyse the proton-photon
correlations in the soft- and hard-photon energy regions.

The paper is organised as follows. In section ~\ref{sec:spectrum} we
write down the correction to the electron--proton energy and angular
distribution of the radiative $\beta^-$--decay of the neutron,
obtained in \cite{Ivanov2012}, which includes the proton--photon
correlations. In section ~\ref{sec:softphoton} we calculate the
contribution of the soft--photon energy region. In section
~\ref{sec:hardphoton} we numerically calculate the contribution of the
hard--photon energy region.  We show that the contributions of the
proton--photon correlations to the lifetime of the neutron $\tau_n$
and the proton recoil energy spectrum $a(T_p)$, integrated over the
proton recoil energy, are of order $10^{-5}$, and can be neglected at
the level of accuracy $10^{-5}$, accepted in \cite{Ivanov2012}. This
confirms the use of the radiative corrections to the lifetime of the
neutron and the proton recoil energy spectrum, described by the
functions $(\alpha/\pi)\,g_n(E_e)$ and $(\alpha/\pi)\,f_n(E_e)$
\cite{Ivanov2012}. In turn the contributions of the proton--photon
correlations to the proton recoil asymmetry $C$ are of order $10^{-4}$
and should be taken into account at the level of accuracy $10^{-5}$
\cite{Ivanov2012}. We show that the account for the proton--photon
correlations to the proton recoil asymmetry $C$ makes the contribution
of the radiative corrections symmetric with respect to a change $A_0
\longleftrightarrow B_0$ as well as the main term $C_0 = - x_C (A_0 +
B_0)$, calculated for the first time by Treiman \cite{Treiman1958},
where $A_0$ and $B_0$ are the correlation coefficients of the neutron
$\beta^-$--decay, calculated to leading order in the large baryon mass
expansion \cite{Abele2008,Nico2009} (see also \cite{Ivanov2012}).  In
section ~\ref{sec:conclusion} we summarise the obtained results. In
section ~\ref{sec:gluck} we compare our results with the results,
obtained by Gl\"uck \cite{Gluck1993}. We show that i) the radiative
corrections to the correlation coefficient $a_0$ in the
electron--antineutrino $(E_e, \cos\theta_{e\bar{\nu}})$ distribution
may be described by the function $f_n(E_e)$ and ii) the radiative
corrections to the correlation coefficient $a_0$ in the proton--energy
spectrum $a(T_p)$, described by the functions $g_n(E_e)$ and
$f_n(E_e)$, are of order of magnitude larger compared with the
radiative corrections from the proton recoil spectrum of the radiative
$\beta^-$--decay of the neutron, caused by the proton--photon
correlations. Since the proton--energy spectrum $a(T_p)$ is obtained
by the integration of the electron--proton energy distribution $a(E_e,
T_p)$ \cite{Ivanov2012} over the electron energies, the results,
concerning the radiative corrections to the proton--energy spectrum
$a(T_p)$, are fully valid for the electron--proton energy distribution
$a(E_e, T_p)$.

\section{Correction to electron--proton energy and angular distribution of 
neutron $\beta^-$--decay accounting for proton-photon correlations}
\label{sec:spectrum}

Using the results, obtained in Appendices B and I of
Ref.\cite{Ivanov2012}, we may write down the correction of the
radiative $\beta^-$--decay of the neutron to the electron--proton
energy and angular distribution of the neutron $\beta^-$--decay
\cite{Ivanov2012}, taking into account the proton-photon
correlations. We get
\begin{eqnarray}\label{label1}
\hspace{-0.3in}&&\frac{d^3\Delta \lambda_{\beta^-_c}(E_e, k_p,
  \theta_p, P)}{dE_e dk_pd\cos\theta_p} = \sum_{j =1,2,3}\frac{d^3
  \Delta \lambda^{(j)}_{\beta^-_c}(E_e, k_p, \theta_p, P)}{dE_e
  dk_pd\cos\theta_p},
\end{eqnarray}
where $E_e$ and $k_p$ are a total electron energy and an absolute
value of the proton 3--momentum, $\theta_p$ is a polar angle between
the proton 3--momentum and the neutron spin and $P$ is the neutron
polarisation \cite{Ivanov2012}. Then, we have denoted
\begin{eqnarray}\label{label2}
\hspace{-0.3in}&&\frac{d^3\Delta \lambda^{(1)}_{\beta^-_c}(E_e, k_p,
  \theta_p, P)}{dE_e dk_pd\cos\theta_p} = (1 +
3\lambda^2)\,\frac{\alpha}{\pi}\,\frac{G^2_F|V_{ud}|^2}{4\pi^3}\,\int
\frac{d\Omega_{ep}}{4\pi}\,\Big\{- \Big(1 - B_0\,\frac{\vec{\xi}_n
  \cdot (\vec{k}_p + \vec{k}_e)}{|\vec{k}_p +
  \vec{k}_e|}\Big)\,g^{(1)}_{\beta^-_c\gamma}(E_e,\mu)\nonumber\\
\hspace{-0.3in}&& - \Big(- a_0\,\frac{\vec{k}_e\cdot (\vec{k}_p +
  \vec{k}_e)}{E_e |\vec{k}_p + \vec{k}_e|} + A_0\,\frac{\vec{\xi}_n \cdot
  \vec{k}_e}{E_e}\Big)\,g^{(2)}_{\beta^-_c\gamma}(E_e,\mu)\Big\}\,\delta(E_0 - E_e - |\vec{k}_p + \vec{k}_e|)\,k_e E_e
k^2_p\,F(E_e, Z = 1),
\end{eqnarray}
\begin{eqnarray}\label{label3}
\hspace{-0.3in}&&\frac{d^3\Delta \lambda^{(2)}_{\beta^-_c}(E_e, k_p,
  \theta_p, P)}{dE_e dk_pd\cos\theta_p} = (1 +
3\lambda^2)\,\frac{\alpha}{\pi}\,\frac{G^2_F|V_{ud}|^2}{4\pi^3}
\int^{2\pi}_0 \frac{d\phi_p}{2\pi} \int \frac{d\Omega_{ep}}{4\pi} \int
\frac{d\omega}{\omega}\int \frac{d\Omega_{\gamma}}{4\pi}\,\Big\{\Big(1
- B_0\,\frac{\vec{\xi}_n \cdot (\vec{k}_p + \vec{k}_e)}{|\vec{k}_p +
  \vec{k}_e|}\Big)\nonumber\\
\hspace{-0.3in}&&\times\,\Big[\frac{k^2_e - (\vec{n}\cdot
  \vec{k}_e)^2}{(E_e - \vec{n}\cdot \vec{k}_e)^2}\,\Big(1 +
\frac{\omega}{E_e}\Big) + \frac{1}{E_e - \vec{n}\cdot
  \vec{k}_e}\,\frac{\omega^2}{E_e}\Big] + \Big(- a_0\,\frac{\vec{k}_p
  + \vec{k}_e }{|\vec{k}_p + \vec{k}_e|} + A_0\,\vec{\xi}_n\Big)\cdot
\Big[\Big(\frac{k^2_e - (\vec{n}\cdot \vec{k}_e)^2}{(E_e -
    \vec{n}\cdot \vec{k}_e)^2} + \frac{\omega}{E_e -\vec{n}\cdot
    \vec{k}_e }\Big)\,\frac{\vec{k}_e}{E_e}\nonumber\\
\hspace{-0.3in}&& + \Big( - \frac{m^2_e}{(E_e - \vec{n}\cdot
  \vec{k}_e)^2} + \frac{E_e + \omega}{E_e - \vec{n}\cdot
  \vec{k}_e}\Big)\,\frac{\omega}{E_e}\,\vec{n}\Big]\Big\}\,\delta(E_0
- E_e - |\vec{k}_p + \vec{k}_e + \omega\,\vec{n}| - \omega)\,k_e E_e
k^2_p\,F(E_e, Z = 1)
\end{eqnarray}
and
\begin{eqnarray}\label{label4}
\hspace{-0.3in}&&\frac{d^3\Delta \lambda^{(3)}_{\beta^-_c}(E_e, k_p,
  \theta_p, P)}{dE_e dk_pd\cos\theta_p} = (1 +
3\lambda^2)\,\frac{\alpha}{\pi}\,\frac{G^2_F|V_{ud}|^2}{4\pi^3}
\int^{2\pi}_0 \frac{d\phi_p}{2\pi}\int \frac{d\Omega_{ep}}{4\pi} \int
\frac{d\omega}{\omega}\int
\frac{d\Omega_{\gamma}}{4\pi}\,\Big\{B_0\,\Big(\frac{\vec{\xi}_n \cdot
  (\vec{k}_p + \vec{k}_e)}{|\vec{k}_p + \vec{k}_e|}\nonumber\\
\hspace{-0.3in}&& - \frac{\vec{\xi}_n \cdot (\vec{k}_p + \vec{k}_e +
  \omega\,\vec{n}\,)}{|\vec{k}_p + \vec{k}_e +
  \omega\,\vec{n}\,|}\Big)\,\Big[\frac{k^2_e - (\vec{n}\cdot
    \vec{k}_e)^2}{(E_e - \vec{n}\cdot \vec{k}_e)^2}\,\Big(1 +
  \frac{\omega}{E_e}\Big) + \frac{1}{E_e - \vec{n}\cdot
    \vec{k}_e}\,\frac{\omega^2}{E_e}\Big] + a_0\,\Big(\frac{ \vec{k}_p
  + \vec{k}_e}{|\vec{k}_p + \vec{k}_e|} - \frac{ \vec{k}_p + \vec{k}_e
  + \omega\,\vec{n}}{|\vec{k}_p + \vec{k}_e +
  \omega\,\vec{n}\,|}\Big)\nonumber\\
\hspace{-0.3in}&&\cdot \Big[\Big(\frac{k^2_e - (\vec{n}\cdot
    \vec{k}_e)^2}{(E_e - \vec{n}\cdot \vec{k}_e)^2} +
  \frac{\omega}{E_e -\vec{n}\cdot \vec{k}_e
  }\Big)\,\frac{\vec{k}_e}{E_e} + \Big( - \frac{m^2_e}{(E_e -
    \vec{n}\cdot \vec{k}_e)^2} + \frac{E_e + \omega}{E_e -
    \vec{n}\cdot
    \vec{k}_e}\Big)\,\frac{\omega}{E_e}\,\vec{n}\Big]\Big\}\nonumber\\
\hspace{-0.3in}&&\times\,\delta(E_0 - E_e - |\vec{k}_p + \vec{k}_e +
\omega\,\vec{n}| - \omega)\,k_e E_e k^2_p\,F(E_e, Z = 1),
\end{eqnarray}
where $\lambda$, $\alpha$, $G_F$ and $V_{ud}$ are the axial,
fine--structure and Fermi coupling constants and the
Cabibbo--Kobayashi--Maskawa (CKM) matrix element, respectively
\cite{Ivanov2012}. Then, $\phi_p$ is an azimuthal angle between the
neutron spin and the proton 3--momentum, $d\Omega_{ep} =
\sin\theta_{ep}d\theta_{ep}d\phi_{ep}$ and $d\Omega_{p\gamma} =
\sin\theta_{p\gamma}d\theta_{p\gamma}d\phi_{p\gamma}$ are the solid
angle elements of the 3--momenta of the electron and photon relative
to a 3--momentum of the proton, $\omega$ and $\omega \vec{n}$ are a
photon energy and a 3--momentum, $\vec{\xi}_n$ is a neutron
polarisation vector such as $\vec{\xi}_n\cdot \vec{k}_p = P
k_p\,\cos\theta_p$ and $\vec{\xi}_n\cdot \vec{k}_e = P
k_p\,(\cos\theta_p \cos\theta_{ep} + \sin\theta_p
\sin\theta_{ep}\,\cos(\phi_p - \phi_{ep}))$ with $P =
|\vec{\xi}_n|$. Then $a_0$, $A_0$ and $B_0$ are the correlation
coefficients of the neutron $\beta^-$--decay, calculated to leading
order in the large baryon mass expansion \cite{Abele2008,Nico2009}
(see also \cite{Ivanov2012}). The functions
$g^{(1)}_{\beta^-_c\gamma}(E_e,\mu)$ and
$g^{(2)}_{\beta^-_c\gamma}(E_e,\mu)$, calculated in Appendix B of
Ref.\cite{Ivanov2012} (see Eq.(B-28)), are equal to
\begin{eqnarray}\label{label5}
\hspace{-0.3in}&&g^{(1)}_{\beta^-_c\gamma}(E_e,\mu) = \Big[{\ell
n}\Big(\frac{2(E_0 - E_e)}{\mu}\Big) - \frac{3}{2} + \frac{1}{3}\,\frac{E_0 -
E_e}{E_e}\, \Big(1 + \frac{1}{8} \frac{E_0 - E_e}{E_e}
\Big)\Big]\Big[\frac{1}{\beta}\,{\ell n}\Big(\frac{1 + \beta}{1 -
    \beta}\Big) - 2\Big] + 1 \nonumber\\
\hspace{-0.3in}&& + \frac{1}{12} \frac{(E_0 - E_e)^2}{E^2_e}+
\frac{1}{2\beta}\,{\ell n}\Big(\frac{1 + \beta}{1 - \beta}\Big) -
\frac{1}{4\beta}\,{\ell n}^2\Big(\frac{1 + \beta}{1 -
  \beta}\Big) + \frac{1}{\beta}\,L\Big(\frac{2 \beta}{1 +
\beta} \Big),\nonumber\\
\hspace{-0.3in}&&g^{(2)}_{\beta^-_c\gamma}(E_e,\mu) = \Big[{\ell
    n}\Big(\frac{2(E_0 - E_e)}{\mu}\Big) - \frac{3}{2} + \frac{1}{3}\,\frac{E_0
    - E_e}{\beta^2 E_e}\, \Big(1 + \frac{1}{8}\,\frac{E_0 -
  E_e}{E_e}\Big) \Big]\Big[\frac{1}{\beta}\,{\ell n}\Big(\frac{1 +
    \beta}{1 - \beta}\Big) - 2\Big] + 1 \nonumber\\
\hspace{-0.3in}&&+ \frac{1}{2\beta}\,{\ell n}\Big(\frac{1 +
\beta}{1 - \beta}\Big) - \frac{1}{4\beta}\,{\ell n}^2\Big(\frac{1 +
\beta}{1 - \beta}\Big) + \frac{1}{\beta}\,L\Big(\frac{2 \beta}{1 +
\beta} \Big),
\end{eqnarray}
where $\mu$ is a finite photon mass or a Lorentz invariant infrared
regularisation scale and $L(x)$ is the Spence function
\cite{HMF72}--\cite{PolyLog3}. The Fermi function $F(E_e,Z=1)$ takes
into account the electron--proton final--state Coulomb interaction
\cite{Ivanov2012}. Having integrated in Eq.(\ref{label2}) over the
solid angle $\Omega_{ep}$ we arrive at the expression
\begin{eqnarray}\label{label6}
\hspace{-0.3in}&&\frac{d^3\Delta \lambda^{(1)}_{\beta^-_c}(E_e, k_p,
  \theta_p, P)}{dE_e dk_pd\cos\theta_p} = (1 +
3\lambda^2)\,\frac{\alpha}{\pi}\,\frac{G^2_F|V_{ud}|^2}{8\pi^3}\,\Big\{-\,
\Big((E_0 - E_e) - P B_0\,\cos\theta_p\,\frac{(E_0 - E_e)^2 + k^2_p - k^2_e}{2 k_p
  }\Big)\,g^{(1)}_{\beta^-_c\gamma}(E_e,\mu)\nonumber\\
\hspace{-0.3in}&& - \,\Big(- a_0\,\frac{(E_0 - E_e)^2 - k^2_p +
  k^2_e}{2 E_e} + P A_0\,\cos\theta_p\,(E_0 - E_e)\,\frac{(E_0 -
  E_e)^2 - k^2_p - k^2_e}{2 k_p
  E_e}\Big)\,g^{(2)}_{\beta^-_c\gamma}(E_e,\mu)\Big\}\,E_e k_p\,F(E_e,
Z = 1),\nonumber\\
\hspace{-0.3in}&&
\end{eqnarray}
where $E = E_0 - E_e$ and $\beta = k_e/E_e = \sqrt{E^2_e - m^2_e}/E_e$
is the electron velocity.  

According to Gl\"uck \cite{Gluck1997}, the calculation of
Eq.(\ref{label3}) and Eq.(\ref{label4}) demands the analysis of the
contributions of the soft- and hard-photon energy regions, where the
contribution of the soft--photon energy region is responsible for a
cancellation of infrared divergences in the energy and angular
distributions of the neutron $\beta^-$--decay. The hard--photon energy
region, being independent of the infrared cut--off, should define a
part of the observable radiative corrections to the lifetime and the
correlation coefficients of the neutron $\beta^-$--decay.

\section{Soft-photon contribution to proton recoil energy and angular 
distribution of neutron radiative $\beta^-$--decay}
\label{sec:softphoton}

In this section we calculate the contributions of the soft--photons,
which are responsible for the cancellation of the infrared divergences
in the neutron $\beta^-$--decay.  For the calculation of
Eq.(\ref{label3}) and Eq.(\ref{label4}) in the soft--photon energy
region we neglect i) correlations between a photon momentum $\vec{q} =
\omega\,\vec{n}$ and a momentum $\vec{k}_p + \vec{k}_e$ and use ii)
for regularisation of infrared divergent contributions a finite-photon
mass regularisation (FPM) \cite{Sirlin1967}--\cite{Shann1971} (see
also \cite{Ivanov2012}). For the application of the FPM to the problem
under consideration we transcribe the right--hand--side (r.h.s.) of
Eq.(\ref{label3}) as follows
\begin{eqnarray*}
\hspace{-0.3in}&&\frac{d^3\Delta \lambda^{(2)}_{\beta^-_c}(E_e, k_p,
  \theta_p, P)}{dE_e dk_pd\cos\theta_p} = (1 +
3\lambda^2)\,\frac{\alpha}{\pi}\,\frac{G^2_F|V_{ud}|^2}{4\pi^3}
\int^{2\pi}_0 \frac{d\phi_p}{2\pi} \int \frac{d\Omega_{ep}}{4\pi} \int
\frac{q^2 dq}{\omega^3}\int \frac{d\Omega_{\gamma}}{4\pi}\,\Big\{(E_0 -
E_e - \omega) - B_0\,\vec{\xi}_n \cdot (\vec{k}_p +
  \vec{k}_e)\nonumber\\
\hspace{-0.3in}&& - a_0\,\frac{(\vec{k}_p + \vec{k}_e)\cdot \vec{k}_e
}{E_e} + A_0\,(E_0 - E_e - \omega)\,\frac{\vec{\xi}_n\cdot
  \vec{k}_e}{E_e}\Big\}\, \frac{k^2_e - (\vec{v}\cdot
  \vec{k}_e)^2}{(E_e - \vec{v}\cdot
  \vec{k}_e)^2}\,\delta\Big(\frac{(E_0 - E_e)(E_0 - E_e - 2\omega) -
  k^2_p - k^2_e}{2 k_e k_p} - \cos\theta_{ep}\Big)\nonumber\\
\hspace{-0.3in}&&\times\,E_e k_p\,F(E_e, Z = 1) + (1 +
3\lambda^2)\,\frac{\alpha}{\pi}\,\frac{G^2_F|V_{ud}|^2}{4\pi^3}
\int^{2\pi}_0 \frac{d\phi_p}{2\pi} \int \frac{d\Omega_{ep}}{4\pi} \int
\frac{d\omega}{E_e}\int \frac{d\Omega_{\gamma}}{4\pi}\,\Big\{\Big((E_0
- E_e - \omega) - B_0\,\vec{\xi}_n \cdot (\vec{k}_p +
\vec{k}_e)\Big)\nonumber\\
\hspace{-0.3in}&&\times\,\Big(\frac{k^2_e - (\vec{n}\cdot
  \vec{k}_e)^2}{(E_e - \vec{n}\cdot \vec{k}_e)^2} + \frac{\omega}{E_e
  - \vec{n}\cdot \vec{k}_e}\Big) + \Big( - a_0\,(\vec{k}_p +
\vec{k}_e) + A_0\,(E_0 - E_e - \omega)\,\vec{\xi}_n\Big)\cdot
\Big(\frac{\vec{k}_e + (E_e + \omega)\,\vec{n}}{E_e -\vec{n}\cdot
  \vec{k}_e } - \frac{m^2_e\,\vec{n}}{(E_e - \vec{n}\cdot
  \vec{k}_e)^2}\Big)\Big\}
\end{eqnarray*}
\begin{eqnarray}\label{label7}
\hspace{-0.3in}&&\times\,\delta\Big(\frac{(E_0 - E_e)(E_0 - E_e -
  2\omega) - k^2_p - k^2_e}{2 k_e k_p} - \cos\theta_{ep}\Big)\,E_e
k_p\,F(E_e, Z = 1),
\end{eqnarray}
where $q = \sqrt{\omega^2 - \mu^2}$ and $v = q/\omega$ are a photon
momentum and velocity, respectively, (see Appendix B of
Ref.\cite{Ivanov2012}).  After the integration over $\phi_p$ and
$\cos\theta_{ep}$ the r.h.s. of Eq.(\ref{label7}) takes the form
\begin{eqnarray}\label{label8}
\hspace{-0.3in}&&\frac{d^3\Delta \lambda^{(2)}_{\beta^-_c}(E_e, k_p,
  \theta_p, P)}{dE_e dk_pd\cos\theta_p} = (1 +
3\lambda^2)\,\frac{\alpha}{\pi}\,\frac{G^2_F|V_{ud}|^2}{8\pi^3} \int
\frac{q^2 dq}{\omega^3}\int \frac{d\Omega_{\gamma}}{4\pi}\, \Big((E_0
- E_e - \omega) - P B_0\,\cos\theta_p\nonumber\\
\hspace{-0.3in}&&\times\,\frac{(E_0 - E_e)(E_0 - E_e - 2 \omega) + k^2_p -
  k^2_e}{2 k_p} - a_0\,\frac{(E_0 - E_e)(E_0 - E_e - 2 \omega) - k^2_p + k^2_e}{2
  E_e } + P A_0\, \cos\theta_p\, (E_0 - E_e - \omega)\nonumber\\
\hspace{-0.3in}&&\times\,\frac{(E_0 - E_e)(E_0 - E_e - 2 \omega) -
  k^2_p - k^2_e}{2k_p E_e}\Big)\, \frac{k^2_e - (\vec{v}\cdot
  \vec{k}_e)^2}{(E_e - \vec{v}\cdot \vec{k}_e)^2}\,E_e k_p\,F(E_e, Z =
1) + (1 +
3\lambda^2)\,\frac{\alpha}{\pi}\,\frac{G^2_F|V_{ud}|^2}{8\pi^3} \int
\frac{d\omega}{E_e}\int \frac{d\Omega_{\gamma}}{4\pi}\nonumber\\
\hspace{-0.3in}&&\times\,\Big\{\Big((E_0 - E_e - \omega) - P
B_0\,\cos\theta_p\,\frac{ (E_0 - E_e)(E_0 - E_e - 2 \omega) + k^2_p -
  k^2_e}{2 k_p }\Big)\,\Big(\frac{k^2_e - (\vec{n}\cdot
  \vec{k}_e)^2}{(E_e - \vec{n}\cdot \vec{k}_e)^2} + \frac{\omega}{E_e
  - \vec{n}\cdot \vec{k}_e}\Big)\nonumber\\
\hspace{-0.3in}&& + \Big( - a_0\,\frac{(E_0 - E_e)(E_0 - E_e - 2
  \omega) - k^2_p + k^2_e}{2} + P A_0\,\cos\theta_p\, (E_0 - E_e -
\omega)\,\frac{(E_0 - E_e)(E_0 - E_e - 2 \omega) -
  k^2_p - k^2_e}{2k_p}\Big)\nonumber\\
\hspace{-0.3in}&&\times\,\frac{1}{k^2_e}\, \Big(\frac{ k^2_e + (E_e +
  \omega)\,(\vec{n}\cdot \vec{k}_e)}{E_e -\vec{n}\cdot \vec{k}_e } -
\frac{m^2_e\,(\vec{n}\cdot \vec{k}_e)}{(E_e - \vec{n}\cdot
  \vec{k}_e)^2}\Big)\Big\}\,E_e k_p\,F(E_e, Z = 1).
\end{eqnarray}
Integrating over directions of a photon velocity $\vec{v}$ and a
momentum $q$ we obtain
\begin{eqnarray}\label{label9}
\hspace{-0.3in}&&\frac{d^3\Delta \lambda^{(2)}_{\beta^-_c}(E_e, k_p,
  \theta_p, P)}{dE_e dk_pd\cos\theta_p} = (1 +
3\lambda^2)\,\frac{\alpha}{\pi}\,\frac{G^2_F|V_{ud}|^2}{8\pi^3}\,
\Big((E_0 - E_e) - P B_0\,\cos\theta_p\,\frac{(E_0 - E_e)^2 + k^2_p -
  k^2_e}{2 k_p} - a_0\nonumber\\
\hspace{-0.3in}&&\times\,\frac{(E_0 - E_e)^2 - k^2_p + k^2_e}{2 E_e }
+ P A_0\, \cos\theta_p\, (E_0 - E_e)\,\frac{(E_0 - E_e)^2 - k^2_p -
  k^2_e}{2k_p E_e}\Big)\,\Big\{{\ell n}\Big(\frac{2(E_0 -
  E_e)}{\mu}\Big)\,\Big[\frac{1}{\beta}\,{\ell n}\Big(\frac{1 +
    \beta}{1 - \beta}\Big) - 2\Big]\nonumber\\
\hspace{-0.3in}&& + 1 + \frac{1}{2\beta}\,{\ell n}\Big(\frac{1 +
  \beta}{1 - \beta}\Big) - \frac{1}{4\beta}\,{\ell n}^2\Big(\frac{1 +
  \beta}{1 - \beta}\Big) + \frac{1}{\beta}\,L\Big(\frac{2\beta}{1 +
  \beta}\Big)\Big\}\,E_e k_p\,F(E_e, Z = 1) + (1 +
3\lambda^2)\,\frac{\alpha}{\pi}\,\frac{G^2_F|V_{ud}|^2}{8\pi^3} \int
\frac{d\omega}{\omega}\int \frac{d\Omega_{\gamma}}{4\pi}\nonumber\\
\hspace{-0.3in}&&\times\, \Big( - \omega - P B_0\,\cos\theta_p\,\frac{
  - 2\omega (E_0 - E_e)}{2 k_p} - a_0\,\frac{ - 2 \omega (E_0 -
  E_e)}{2 E_e } + P A_0\, \cos\theta_p\, (- \omega)\,\frac{(E_0 -
  E_e)(E_0 - E_e - 2 \omega) - k^2_p - k^2_e}{2k_p E_e}\nonumber\\
\hspace{-0.3in}&& + P A_0\cos\theta_p\,(E_0 - E_e)\,\frac{ - 2 \omega
  (E_0 - E_e) }{2k_p E_e}\Big)\, \frac{k^2_e - (\vec{n}\cdot
  \vec{k}_e)^2}{(E_e - \vec{n}\cdot \vec{k}_e)^2}\,E_e k_p\,F(E_e, Z =
1) + (1 +
3\lambda^2)\,\frac{\alpha}{\pi}\,\frac{G^2_F|V_{ud}|^2}{8\pi^3} \int
\frac{d\omega}{E_e}\int \frac{d\Omega_{\gamma}}{4\pi}\nonumber\\
\hspace{-0.3in}&&\times\,\Big\{\Big((E_0 - E_e - \omega) - P
B_0\,\cos\theta_p\,\frac{(E_0 - E_e)(E_0 - E_e - 2 \omega) + k^2_p -
  k^2_e}{2 k_p }\Big)\Big(\frac{k^2_e - (\vec{n}\cdot
  \vec{k}_e)^2}{(E_e - \vec{n}\cdot \vec{k}_e)^2} + \frac{\omega}{E_e
  - \vec{n}\cdot \vec{k}_e}\Big)\nonumber\\
\hspace{-0.3in}&& + \Big( - a_0\,\frac{(E_0 - E_e)(E_0 - E_e - 2
  \omega) - k^2_p + k^2_e}{2} + P A_0\,\cos\theta_p\, (E_0 - E_e -
\omega)\frac{(E_0 - E_e)(E_0 - E_e - 2 \omega) -
  k^2_p - k^2_e}{2k_p}\Big)\nonumber\\
\hspace{-0.3in}&&\times\,\frac{1}{k^2_e}\, \Big(\frac{ k^2_e + (E_e +
  \omega)\,(\vec{n}\cdot \vec{k}_e)}{E_e -\vec{n}\cdot \vec{k}_e } -
\frac{m^2_e\,(\vec{n}\cdot \vec{k}_e)}{(E_e - \vec{n}\cdot
  \vec{k}_e)^2}\Big)\Big\}\,E_e k_p\,F(E_e, Z = 1).
\end{eqnarray}
Summing up the contributions of Eqs.(\ref{label6}) and (\ref{label9})
we obtain the expression, which does not depend on a photon mass or an
infrared regularisation scale $\mu$, that is,
\begin{eqnarray*}
\hspace{-0.3in}&&\sum_{j = 1,2}\frac{d^3\Delta
  \lambda^{(j)}_{\beta^-_c}(E_e, k_p, \theta_p, P)}{dE_e
  dk_pd\cos\theta_p} = (1 +
3\lambda^2)\,\frac{\alpha}{\pi}\,\frac{G^2_F|V_{ud}|^2}{8\pi^3}\,
\Big\{\Big((E_0 - E_e) - P B_0\,\cos\theta_p\,\frac{(E_0 - E_e)^2 +
  k^2_p - k^2_e}{2 k_p}\Big)\nonumber\\
\hspace{-0.3in}&&\times\,\Big[\Big(\frac{3}{2} -
  \frac{1}{3}\,\frac{(E_0 - E_e)}{E_e} - \frac{1}{24}\,\frac{(E_0 -
    E_e)^2}{E^2_e}\Big)\Big[\frac{1}{\beta}\,{\ell n}\Big(\frac{1 +
      \beta}{1 - \beta}\Big) - 2\Big] - \frac{1}{12}\,\frac{(E_0 -
    E_e)^2}{E^2_e}\Big] + \Big( - a_0\,\frac{(E_0 - E_e)^2 - k^2_p +
  k^2_e}{2 E_e }\nonumber\\
\hspace{-0.3in}&& + P A_0\, \cos\theta_p\, (E_0 - E_e)\,\frac{(E_0 -
  E_e)^2 - k^2_p - k^2_e}{2k_p E_e}\Big)\,\Big(\frac{3}{2} -
  \frac{1}{3}\,\frac{(E_0 - E_e)}{\beta^2 E_e} -
  \frac{1}{24}\,\frac{(E_0 - E_e)^2}{\beta^2
    E^2_e}\Big)\Big[\frac{1}{\beta}\,{\ell n}\Big(\frac{1 + \beta}{1 -
      \beta}\Big) - 2\Big]\Big\}\nonumber\\
\hspace{-0.3in}&&\times\,E_e k_p\,F(E_e, Z = 1) \nonumber\\
\end{eqnarray*}
\vspace{-0.3in}
\begin{eqnarray*}
\hspace{-0.3in}&&
 + (1 +
3\lambda^2)\,\frac{\alpha}{\pi}\,\frac{G^2_F|V_{ud}|^2}{8\pi^3} \int
\frac{d\Omega_{\gamma}}{4\pi}\,\int d\omega\,\Big\{ - 1 - P
B_0\,\cos\theta_p\,\frac{ - 2 (E_0 - E_e)}{2 k_p} - a_0\,\frac{ - 2
  (E_0 - E_e)}{2 E_e }\nonumber\\
\hspace{-0.3in}&& + P A_0\, \cos\theta_p\, (- 1)\,\frac{(E_0 -
  E_e)(E_0 - E_e - \omega) - k^2_p - k^2_e}{2k_p E_e} + P
A_0\cos\theta_p\,(E_0 - E_e)\,\frac{ - 2 (E_0 - E_e)}{2k_p
  E_e}\Big\}\, \frac{k^2_e - (\vec{n}\cdot \vec{k}_e)^2}{(E_e -
  \vec{n}\cdot \vec{k}_e)^2}\nonumber\\
\hspace{-0.3in}&&\times\,E_e k_p\,F(E_e, Z = 1)\nonumber\\
\end{eqnarray*}
\vspace{-0.5in}
\begin{eqnarray}\label{label10}
\hspace{-0.3in}&& + (1 +
3\lambda^2)\,\frac{\alpha}{\pi}\,\frac{G^2_F|V_{ud}|^2}{8\pi^3} \int
\frac{d\omega}{E_e}\int \frac{d\Omega_{\gamma}}{4\pi}\,\Big\{\Big((E_0
- E_e - \omega) - P B_0\,\cos\theta_p\,\frac{(E_0 - E_e)(E_0 - E_e - 2
  \omega) + k^2_p - k^2_e}{2 k_p }\Big)\nonumber\\
\hspace{-0.3in}&&\times\,\Big(\frac{k^2_e - (\vec{n}\cdot
  \vec{k}_e)^2}{(E_e - \vec{n}\cdot \vec{k}_e)^2} + \frac{\omega}{E_e
  - \vec{n}\cdot \vec{k}_e}\Big) + \Big( - a_0\,\frac{(E_0 - E_e)(E_0
  - E_e - 2 \omega) - k^2_p + k^2_e}{2} + P A_0\,\cos\theta_p\, (E_0 -
E_e - \omega)\nonumber\\
\hspace{-0.3in}&&\times\,\frac{(E_0 - E_e)(E_0 - E_e - 2\omega) -
  k^2_p - k^2_e}{2k_p}\Big)\,\frac{1}{k^2_e}\, \Big(\frac{ k^2_e +
  (E_e + \omega)\,(\vec{n}\cdot \vec{k}_e)}{E_e -\vec{n}\cdot
  \vec{k}_e } - \frac{m^2_e\,(\vec{n}\cdot \vec{k}_e)}{(E_e -
  \vec{n}\cdot \vec{k}_e)^2}\Big)\Big\}\,E_e k_p\,F(E_e, Z = 1).
\end{eqnarray}
Having integrated Eq.(\ref{label10}) over $\omega$ in the limits $0
\le \omega \le (E_0 - E_e)$ and directions of a photon momentum we
find the expression
\begin{eqnarray}\label{label11}
\hspace{-0.3in}&&\sum_{j = 1,2}\frac{d^3\Delta
  \lambda^{(j)}_{\beta^-_c}(E_e, k_p, \theta_p, P)}{dE_e
  dk_pd\cos\theta_p} = (1 +
3\lambda^2)\,\frac{\alpha}{\pi}\,\frac{G^2_F|V_{ud}|^2}{8\pi^3}\,
\Big\{\Big((E_0 - E_e) - P B_0\,\cos\theta_p\,\frac{(E_0 - E_e)^2 +
  k^2_p - k^2_e}{2 k_p}\Big)\nonumber\\
\hspace{-0.3in}&&\times\,\Big\{\Big(\frac{3}{2} -
  \frac{1}{3}\,\frac{(E_0 - E_e)}{E_e} - \frac{1}{24}\,\frac{(E_0 -
    E_e)^2}{E^2_e}\Big)\Big[\frac{1}{\beta}\,{\ell n}\Big(\frac{1 +
      \beta}{1 - \beta}\Big) - 2\Big] - \frac{1}{12}\,\frac{(E_0 -
    E_e)^2}{E^2_e}\Big\} + \Big( - a_0\,\frac{(E_0 - E_e)^2 - k^2_p +
  k^2_e}{2 E_e }\nonumber\\
\hspace{-0.3in}&& + P A_0\, \cos\theta_p\, (E_0 - E_e)\,\frac{(E_0 -
  E_e)^2 - k^2_p - k^2_e}{2k_p E_e}\Big)\,\Big(\frac{3}{2} -
  \frac{1}{3}\,\frac{(E_0 - E_e)}{\beta^2 E_e} -
  \frac{1}{24}\,\frac{(E_0 - E_e)^2}{\beta^2
    E^2_e}\Big)\, \Big[\frac{1}{\beta}\,{\ell n}\Big(\frac{1 + \beta}{1 -
      \beta}\Big) - 2\Big]\Big\}\nonumber\\
\hspace{-0.3in}&&\times\,E_e k_p\,F(E_e, Z = 1)\nonumber\\
\hspace{-0.3in}&&+ (1 +
3\lambda^2)\,\frac{\alpha}{\pi}\,\frac{G^2_F|V_{ud}|^2}{8\pi^3}\,\Big\{
- (E_0 - E_e) + P B_0 \,\cos\theta_p\,\frac{(E_0 -
  E_e)^2}{k_p} + a_0 \,\frac{(E_0 - E_e)^2}{E_e}\nonumber\\
\hspace{-0.3in}&& - P A_0\,\cos\theta_p\,\frac{1}{4}\,\frac{5 (E_0 -
  E_e)^3 - 2 (k^2_p + k^2_e)(E_0 - E_e)}{k_p
  E_e}\Big\}\,\Big[\frac{1}{\beta}\,{\ell n}\Big(\frac{1 + \beta}{1 -
    \beta}\Big) - 2\Big]\,E_e k_p\,F(E_e, Z = 1) \nonumber\\
\hspace{-0.3in}&&+ (1 +
3\lambda^2)\,\frac{\alpha}{\pi}\,\frac{G^2_F|V_{ud}|^2}{8\pi^3}\,\Big\{
\Big(\frac{1}{2}\, \frac{(E_0 - E_e)^2}{E_e} - P
  B_0\,\cos\theta_p\,\frac{1}{2}\,\frac{(k^2_p - k^2_e)\,(E_0 -
    E_e)}{k_p E_e}\Big)\, \Big[\frac{1}{\beta}\,{\ell n}\Big(\frac{1 +
      \beta}{1 - \beta}\Big) - 2\Big]\nonumber\\
\hspace{-0.3in}&& + \Big(\frac{1}{12}\, \frac{(E_0 - E_e)^3}{E^2_e} +
P B_0\,\cos\theta_p\,\frac{1}{24}\,\frac{(E_0 - E_e)^4 - 3 (k^2_p -
  k^2_e)\,(E_0 - E_e)^2}{k_p E^2_e}\Big)\,\frac{1}{\beta}\,{\ell
  n}\Big(\frac{1 + \beta}{1 - \beta}\Big)\nonumber\\
\hspace{-0.3in}&& + \Big(a_0\,\frac{1}{2}\,\frac{ (k^2_p -
  k^2_e)\,(E_0 - E_e)}{E^2_e} + P
A_0\,\cos\theta_p\,\frac{1}{12}\,\frac{(E_0 - E_e)^4 - 3 (k^2_p +
  k^2_e)\,(E_0 - E_e)^2}{k_p
  E^2_e}\Big)\,\frac{1}{\beta^2}\,\Big[\frac{1}{\beta}\,{\ell
    n}\Big(\frac{1 + \beta}{1 - \beta}\Big) - 2\Big]\nonumber\\
\hspace{-0.3in}&& + \Big(a_0\,\frac{1}{24}\,\frac{(E_0 - E_e)^4 + 3
  (k^2_p - k^2_e)\,(E_0 - E_e)^2}{E^3_e} - P
A_0\,\cos\theta_p\,\frac{1}{24}\,\frac{(k^2_p + k^2_e)\,(E_0 -
  E_e)^3}{k_p
  E^3_e}\Big)\,\frac{1}{\beta^2}\,\Big[\frac{1}{\beta}\,{\ell
    n}\Big(\frac{1 + \beta}{1 - \beta}\Big) - 2\Big]\Big\}\nonumber\\
\hspace{-0.3in}&&\times\,E_e
k_p\,F(E_e, Z = 1).
\end{eqnarray}
For the integration over directions of a photon momentum $\vec{q} =
\omega\,\vec{n}$ we have used the formulas
\begin{eqnarray}\label{label12}
\hspace{-0.3in}&&\int \frac{d\Omega_{\gamma}}{4\pi}\,\frac{k^2_e -
  (\vec{n}\cdot \vec{k}_e)^2}{(E_e - \vec{n}\cdot \vec{k}_e)^2} =
\frac{1}{\beta}\,{\ell n}\Big(\frac{1 + \beta}{1 - \beta}\Big) - 2
\;,\; \int \frac{d\Omega_{\gamma}}{4\pi}\, \frac{1}{k^2_e}\,\frac{
  k^2_e + E_e\,(\vec{n}\cdot \vec{k}_e)}{E_e -\vec{n}\cdot \vec{k}_e }
= \frac{1}{2 E_e \beta^2}\Big[\frac{1 + \beta^2}{\beta}\,{\ell
    n}\Big(\frac{1 + \beta}{1 - \beta}\Big) - 2\Big],\nonumber\\
\hspace{-0.3in}&&\int \frac{d\Omega_{\gamma}}{4\pi}\,\frac{1}{E_e - \vec{n}\cdot
  \vec{k}_e} = \frac{1}{2 E_e \beta}\,{\ell n}\Big(\frac{1 + \beta}{1
  - \beta}\Big)\;,\;\int
\frac{d\Omega_{\gamma}}{4\pi}\,\frac{m^2_e}{k^2_e}\,\frac{(\vec{n}\cdot
\vec{k}_e)}{(E_e - \vec{n}\cdot \vec{k}_e)^2} = - \frac{1}{2 E_e
       \beta^2}\Big[\frac{1 - \beta^2}{\beta}\,{\ell n}\Big(\frac{1 +
         \beta}{1 - \beta}\Big) - 2\Big],\nonumber\\
\hspace{-0.3in}&&\int
\frac{d\Omega_{\gamma}}{4\pi}\,\frac{1}{k^2_e}\,\frac{(\vec{n}\cdot
  \vec{k}_e)}{E_e - \vec{n}\cdot \vec{k}_e} = \frac{1}{2 E^2_e
  \beta^2}\Big[\frac{1}{\beta}\,{\ell n}\Big(\frac{1 +
    \beta}{1 - \beta}\Big) - 2\Big].
\end{eqnarray}
Following the approximation, neglecting correlations between a photon
momentum $\vec{q} = \omega\,\vec{n}$ and a momentum $\vec{k}_p +
\vec{k}_e$, for the correction to the electron--proton energy and
angular distribution of the neutron $\beta^-$--decay, given by
Eq.(\ref{label4}), we obtain the expression
\begin{eqnarray}\label{label13}
\hspace{-0.3in}&&\frac{d^3\Delta \lambda^{(3)}_{\beta^-_c}(E_e, k_p,
  \theta_p, P)}{dE_e dk_pd\cos\theta_p} = (1 +
3\lambda^2)\,\frac{\alpha}{\pi}\,\frac{G^2_F|V_{ud}|^2}{8\pi^3} \int
\frac{d\Omega_{\gamma}}{4\pi}\,\Big\{P
B_0\,\cos\theta_p\,\Big[\frac{(k^2_p + k^2_e)(E_0 - E_e)}{2
    k_p}\,\frac{\vec{n}\cdot \vec{k}_e}{k^2_e}\,\frac{k^2_e -
    (\vec{n}\cdot \vec{k}_e)^2}{(E_e - \vec{n}\cdot \vec{k}_e)^2}\nonumber\\
\hspace{-0.3in}&& - \frac{1}{12}\,\frac{(E_0 - E_e)^4 + 3 (k^2_p +
  k^2_e)(E_0 - E_e)^2}{k_p E_e}\,\frac{\vec{n}\cdot
  \vec{k}_e}{k^2_e}\,\frac{k^2_e - (\vec{n}\cdot \vec{k}_e)^2}{(E_e -
  \vec{n}\cdot \vec{k}_e)^2} - \frac{1}{12}\,\frac{(E_0 - E_e)^5 + 2
  (k^2_p + k^2_e)(E_0 - E_e)^3}{k_p E_e}\nonumber\\
\hspace{-0.3in}&&\times\,\frac{\vec{n}\cdot
  \vec{k}_e}{k^2_e}\,\frac{1}{E_e - \vec{n}\cdot \vec{k}_e}\Big] -
a_0\Big[(E_0 - E_e)\frac{\vec{n}\cdot \vec{k}_e}{E_e}\,\frac{k^2_e -
    (\vec{n}\cdot \vec{k}_e)^2}{(E_e - \vec{n}\cdot \vec{k}_e)^2} +
  \frac{1}{2}\,\frac{(E_0 - E_e)^2}{E_e}\,\Big(\frac{\vec{n}\cdot
    \vec{k}_e}{E_e - \vec{n}\cdot \vec{k}_e} - \frac{m^2_e}{(E_e -
    \vec{n}\cdot \vec{k}_e)^2}\nonumber\\
\hspace{-0.3in}&& + \frac{E_e}{E_e - \vec{n}\cdot \vec{k}_e}\Big) +
\frac{1}{3}\,\frac{(E_0 - E_e)^3}{E_e}\,\frac{1}{E_e - \vec{n}\cdot
  \vec{k}_e}\Big]\Big\}\,E_e k_p\,F(E_e, Z = 1).
\end{eqnarray}
The integrals over directions of a photon momentum $\vec{q} =
\omega\,\vec{n}$ are equal to
\begin{eqnarray}\label{label14}
\hspace{-0.3in}&& \int \frac{d\Omega_{\gamma}}{4\pi}\,\frac{\vec{n}
  \cdot \vec{k}_e}{k^2_e}\,\frac{k^2_e - (\vec{n}\cdot
  \vec{k}_e)^2}{(E_e - \vec{n}\cdot \vec{k}_e)^2} =
\frac{1}{2 E_e\beta^2}\,\Big[\frac{3 - \beta^2}{\beta}\,{\ell
    n}\Big(\frac{1 + \beta}{1 - \beta}\Big) -
  6\Big]\;,\;\int \frac{d\Omega_{\gamma}}{4\pi}\,\frac{1}{E_e -
  \vec{n}\cdot \vec{k}_e} = \frac{1}{2 E_e\beta}\,{\ell n}\Big(\frac{1
  + \beta}{1 - \beta}\Big),\nonumber\\
\hspace{-0.3in}&& \int
\frac{d\Omega_{\gamma}}{4\pi}\,\frac{(\vec{n}\cdot
  \vec{k}_e)}{E_e}\,\frac{k^2_e - (\vec{n}\cdot \vec{k}_e)^2}{(E_e -
  \vec{n}\cdot \vec{k}_e)^2} = \frac{1}{2}\,\Big[\frac{3 -
    \beta^2}{\beta}\,{\ell n}\Big(\frac{1 + \beta}{1 - \beta}\Big) -
  6\Big]\;,\; \int
\frac{d\Omega_{\gamma}}{4\pi}\,\frac{m^2_e}{(E_e - \vec{n}\cdot
  \vec{k}_e)^2} = 1,\nonumber\\
\hspace{-0.3in}&& \int \frac{d\Omega_{\gamma}}{4\pi}\,\frac{\vec{n}
  \cdot \vec{k}_e}{k^2_e}\,\frac{1}{E_e - \vec{n}\cdot \vec{k}_e} =
\frac{1}{2 E^2_e\beta^2}\,\Big[\frac{1}{\beta}\,{\ell n}\Big(\frac{1 +
    \beta}{1 - \beta}\Big) - 2\Big]\;,\; \int
\frac{d\Omega_{\gamma}}{4\pi}\,\frac{\vec{n}\cdot
  \vec{k}_e}{E_e}\,\frac{1}{E_e - \vec{n}\cdot \vec{k}_e} = \frac{1}{2
  E_e}\,\Big[\frac{1}{\beta}\,{\ell n}\Big(\frac{1 + \beta}{1 -
    \beta}\Big) - 2\Big].
\end{eqnarray}
Substituting Eq.(\ref{label14}) into Eq.(\ref{label13}) we obtain the
following contribution of Eq.(\ref{label4}) to the electron--proton
energy and angular distribution of the neutron $\beta^-$--decay:
\begin{eqnarray}\label{label15}
\hspace{-0.3in}&&\frac{d^3\Delta \lambda^{(3)}_{\beta^-_c}(E_e, k_p,
  \theta_p, P)}{dE_e dk_pd\cos\theta_p} = (1 +
3\lambda^2)\,\frac{\alpha}{\pi}\,\frac{G^2_F|V_{ud}|^2}{16\pi^3}
\,\Bigg\{ P B_0\,\cos\theta_p\,\Big\{\Big(- \frac{1}{12}\,\frac{(E_0 -
  E_e)^4 + 3 (k^2_p + k^2_e)(E_0 - E_e)^2}{ \beta^2 k_p
  E^2_e}\nonumber\\
\hspace{-0.3in}&& + \frac{1}{2}\,\frac{(k^2_p + k^2_e)(E_0 -
  E_e)}{\beta^2 k_p E_e}\Big)\,\Big[\frac{3 - \beta^2}{\beta}\,{\ell
    n}\Big(\frac{1 + \beta}{1 - \beta}\Big) - 6\Big] -
\frac{1}{12}\,\frac{(E_0 - E_e)^5 + 2 (k^2_p + k^2_e)(E_0 -
  E_e)^3}{\beta^2 k_p E^3_e}\,\Big[\frac{1}{\beta}\,{\ell n}\Big(\frac{1 +
    \beta}{1 - \beta}\Big) - 2\Big]\Big\}\nonumber\\
\hspace{-0.3in}&& - a_0\, \Big\{(E_0 - E_e)\,\Big[\frac{3 -
    \beta^2}{\beta}\,{\ell n}\Big(\frac{1 + \beta}{1 - \beta}\Big) -
  6\Big]+ \frac{(E_0 - E_e)^2}{E_e}\,\Big[\frac{1}{\beta}\,{\ell
    n}\Big(\frac{1 + \beta}{1 - \beta}\Big) - 2\Big] +
\frac{1}{3}\,\frac{(E_0 - E_e)^3}{E^2_e}\, \frac{1}{\beta}\,{\ell
  n}\Big(\frac{1 + \beta}{1 - \beta}\Big)\Big\}\Bigg\}\nonumber\\
\hspace{-0.3in}&&\times\,E_e k_p\,F(E_e, Z = 1).
\end{eqnarray}
Summing up the contributions of Eq.(\ref{label11}) and
Eq.(\ref{label15}) we obtain the total correction to the
electron--proton energy and angular distribution of the neutron
$\beta^-$--decay from the soft--photon energy region. Integrating over
the electron energy $E_e$ and the proton momentum $k_p$ we define a
correction to the proton recoil angular distribution, related to the
proton recoil asymmetry $C$ \cite{Ivanov2012}. We get
\begin{eqnarray}\label{label16}
\hspace{-0.3in}&&\frac{d\Delta \lambda_{\beta^-_c}( \theta_p,
  P)}{d\cos\theta_p} = (1 +
3\lambda^2)\,\frac{G^2_F|V_{ud}|^2}{16\pi^3}
\,\Big\{\frac{\alpha}{\pi}\,\Delta X^{(s)}_2 +
a_0\,\frac{\alpha}{\pi}\,\Delta X^{(s)}_0 + P
\cos\theta_p\,\Big(A_0\,\frac{\alpha}{\pi}\,\Delta X^{(s)}_{10} -
B_0\,\frac{\alpha}{\pi}\,\Delta X^{(s)}_{11}\Big)\Big\},
\end{eqnarray}
where $\Delta X^{(s)}_2$, $\Delta X^{(s)}_0$, $\Delta X^{(s)}_{10}$
and $\Delta X^{(s)}_{11}$ are given by
\begin{eqnarray}\label{label17}
\hspace{-0.3in}&&\Delta X^{(s)}_2 = \int^{k_2}_{k_1}dk_p\int^{(E_e)_{\rm
    max}}_{E_m}dE_e (E_0 - E_e)\,\Big\{\Big(1
+\frac{1}{3}\,\frac{(E_0 - E_e)}{E_e}+ \frac{1}{12}\,\frac{(E_0 -
  E_e)^2}{E^2_e}\Big)\,\Big[\frac{1}{\beta}\,{\ell n}\Big(\frac{1 +
    \beta}{1 - \beta}\Big) - 2\Big]\nonumber\\
\hspace{-0.3in}&& + \frac{1}{6}\,\frac{(E_0 -
  E_e)^2}{E^2_e}\Big\}\,E_e k_p F(E_e, Z = 1) +
\int^{(k_p)_{\rm max}}_{k_2}dk_p\int^{(E_e)_{\rm max}}_{(E_e)_{\rm
    min}} dE_e (E_0 - E_e)\,\Big\{\Big(1 +\frac{1}{3}\,\frac{(E_0 -
  E_e)}{E_e}\nonumber\\
\hspace{-0.3in}&& + \frac{1}{12}\,\frac{(E_0 -
  E_e)^2}{E^2_e}\Big)\,\Big[\frac{1}{\beta}\,{\ell n}\Big(\frac{1 +
    \beta}{1 - \beta}\Big) - 2\Big] + \frac{1}{6}\,\frac{(E_0 -
  E_e)^2}{E^2_e}\Big\}\,E_e k_p F(E_e, Z = 1) = 0.016051,
\,{\rm MeV^5}
\end{eqnarray}
\begin{eqnarray*}
\hspace{-0.3in}&&\Delta X^{(s)}_0 = \int^{k_2}_{k_1}dk_p\int^{(E_e)_{\rm
    max}}_{E_m}dE_e \Bigg\{\Big\{(E_0 - E_e)^2 - ((E_0 - E_e)^2
- k^2_p + k^2_e)\nonumber\\
\hspace{-0.3in}&&\times\,\Big(\frac{3}{2} - \frac{1}{3}\,\frac{(E_0 -
  E_e)}{\beta^2 E_e} - \frac{1}{24}\,\frac{(E_0 - E_e)^2}{\beta^2
  E^2_e}\Big) + \frac{(k^2_p - k^2_e)(E_0 - E_e)}{\beta^2 E_e} +
\frac{1}{12}\,\frac{(E_0 - E_e)^4 + 3 (k^2_p - k^2_e)(E_0 -
  E_e)^2}{\beta^2 E^2_e}\Big\}
\end{eqnarray*}
\begin{eqnarray}\label{label18}
\hspace{-0.3in}&&\times\,\Big[\frac{1}{\beta}\,{\ell n}\Big(\frac{1 +
    \beta}{1 - \beta}\Big) - 2\Big] - \frac{1}{3}\,\frac{(E_0 -
  E_e)^3}{E_e}\,\frac{1}{\beta}\,{\ell n}\Big(\frac{1 + \beta}{1 -
  \beta}\Big)- E_e (E_0 - E_e)\,\Big[\frac{3 -
    \beta^2}{\beta}\,{\ell n}\Big(\frac{1 + \beta}{1 - \beta}\Big) -
  6\Big] \Bigg\}\nonumber\\
\hspace{-0.3in}&& \times\,k_p F(E_e, Z = 1) +
\int^{(k_p)_{\rm max}}_{k_2}dk_p\int^{(E_e)_{\rm max}}_{(E_e)_{\rm
    min}}dE_e \Bigg\{\Big\{(E_0 - E_e)^2 - ((E_0 - E_e)^2 - k^2_p
+ k^2_e)\nonumber\\
\hspace{-0.3in}&&\times\,\Big(\frac{3}{2} - \frac{1}{3}\,\frac{(E_0 -
  E_e)}{\beta^2 E_e} - \frac{1}{24}\,\frac{(E_0 - E_e)^2}{\beta^2
  E^2_e}\Big) + \frac{(k^2_p - k^2_e)(E_0 - E_e)}{\beta^2 E_e} +
\frac{1}{12}\,\frac{(E_0 - E_e)^4 + 3 (k^2_p - k^2_e)(E_0 -
  E_e)^2}{\beta^2 E^2_e}\Big\}\nonumber\\
\hspace{-0.3in}&&\times\,\Big[\frac{1}{\beta}\,{\ell n}\Big(\frac{1 +
    \beta}{1 - \beta}\Big) - 2\Big] - \frac{1}{3}\,\frac{(E_0 -
  E_e)^3}{E_e}\,\frac{1}{\beta}\,{\ell n}\Big(\frac{1 + \beta}{1 -
  \beta}\Big)- E_e (E_0 - E_e)\,\Big[\frac{3 -
    \beta^2}{\beta}\,{\ell n}\Big(\frac{1 + \beta}{1 - \beta}\Big) -
  6\Big] \Bigg\}\nonumber\\
\hspace{-0.3in}&& \times\,k_p F(E_e, Z = 1) = -
0.010201\,{\rm MeV^5},
\end{eqnarray}
\begin{eqnarray}\label{label19}
\hspace{-0.3in}&&\Delta X^{(s)}_{10} =
\int^{k_2}_{k_1}dk_p\int^{(E_e)_{\rm max}}_{E_m}dE_e\Big\{\Big((E_0 -
E_e)^3 - (k^2_p + k^2_e)(E_0 - E_e)\Big)\, \Big(\frac{3}{2} -
\frac{1}{3}\,\frac{(E_0 - E_e)}{\beta^2 E_e} -
\frac{1}{24}\,\frac{(E_0 - E_e)^2}{\beta^2 E^2_e}\Big)\nonumber\\
\hspace{-0.3in}&& - \frac{1}{2}\,(5 (E_0 - E_e)^3 - 2 (k^2_p +
k^2_e)(E_0 - E_e))- \frac{1}{12}\,\frac{(k^2_p + k^2_e)(E_0 -
  E_e)^3}{\beta^2 E^2_e} + \frac{1}{6}\,\frac{(E_0 - E_e)^4 - 3(k^2_p
  + k^2_e)(E_0 - E_e)^2}{\beta^2 E_e}\Big\}\nonumber\\
\hspace{-0.3in}&&\times\,\Big[\frac{1}{\beta}\,{\ell n}\Big(\frac{1 +
    \beta}{1 - \beta}\Big) - 2\Big]\, F(E_e, Z = 1) +
\int^{(k_p)_{\rm max}}_{k_2}dk_p \int^{(E_e)_{\rm max}}_{(E_e)_{\rm
    min}}dE_e \Big\{\Big((E_0 -
E_e)^3 - (k^2_p + k^2_e)(E_0 - E_e)\Big)\nonumber\\
\hspace{-0.3in}&&\times\,\Big(\frac{3}{2} - \frac{1}{3}\,\frac{(E_0 -
  E_e)}{\beta^2 E_e} - \frac{1}{24}\,\frac{(E_0 - E_e)^2}{\beta^2
  E^2_e}\Big) - \frac{1}{2}\,(5 (E_0 - E_e)^3 - 2 (k^2_p + k^2_e)(E_0
- E_e)) - \frac{1}{12}\,\frac{(k^2_p + k^2_e)(E_0 - E_e)^3}{\beta^2
  E^2_e}\nonumber\\
\hspace{-0.3in}&& + \frac{1}{6}\,\frac{(E_0 - E_e)^4 - 3(k^2_p +
  k^2_e)(E_0 - E_e)^2}{\beta^2 E_e}\Big\}\,\Big[\frac{1}{\beta}\,{\ell
    n}\Big(\frac{1 + \beta}{1 - \beta}\Big) - 2\Big]\, F(E_e, Z = 1) =
- 0.045245\,{\rm MeV^5}
\end{eqnarray}
and
\begin{eqnarray}\label{label20}
\hspace{-0.3in}&&\Delta X^{(s)}_{11} =
\int^{k_2}_{k_1}dk_p\int^{(E_e)_{\rm max}}_{E_m}dE_e \Bigg\{E_e
\Big((E_0 - E_e)^2 + k^2_p - k^2_e\Big)\, \Big\{\Big(\frac{3}{2} -
\frac{1}{3}\,\frac{(E_0 - E_e)}{E_e} - \frac{1}{24}\,\frac{(E_0 -
  E_e)^2}{ E^2_e}\Big)\nonumber\\
\hspace{-0.3in}&& \times\,\Big[\frac{1}{\beta}\,{\ell n}\Big(\frac{1 +
    \beta}{1 - \beta}\Big) - 2\Big] - \frac{1}{12}\,\frac{(E_0 - E_e)^2}{ E^2_e}\Big\} -
\Big\{2\,E_e (E_0 - E_e)^2-
\frac{1}{6}\,\frac{(E_0 - E_e)^5 + 2 (k^2_p + k^2_e)(E_0 -
  E_e)^3}{\beta^2 E^2_e}\nonumber\\
\hspace{-0.3in}&&- (k^2_p - k^2_e)\,(E_0 - E_e)
\Big\}\,\Big[\frac{1}{\beta}\,{\ell n}\Big(\frac{1 + \beta}{1 -
    \beta}\Big) - 2\Big] - \frac{1}{12}\,\frac{(E_0 - E_e)^4 - 3 (k^2_p -
  k^2_e)\,(E_0 - E_e)^2}{E_e}\,\frac{1}{\beta}\,{\ell n}\Big(\frac{1 +
  \beta}{1 - \beta}\Big)\nonumber\\
\hspace{-0.3in}&& + \Big\{\frac{1}{6}\,\frac{((E_0 - E_e)^4 + 3 (k^2_p
  + k^2_e)(E_0 - E_e)^2}{\beta^2 E_e} - \frac{(k^2_p + k^2_e)(E_0 -
  E_e)}{\beta^2 }\Big\}\,\Big[\frac{3 - \beta^2}{\beta}\,{\ell
    n}\Big(\frac{1 + \beta}{1 - \beta}\Big) - 6\Big]\Bigg\}\,F(E_e, Z
= 1) \nonumber\\
\hspace{-0.3in}&& + \int^{(k_p)_{\rm max}}_{k_2}dk_p\int^{(E_e)_{\rm
    max}}_{(E_e)_{\rm min}}dE_e\Bigg\{E_e
\Big((E_0 - E_e)^2 + k^2_p - k^2_e\Big)\, \Big\{\Big(\frac{3}{2} -
\frac{1}{3}\,\frac{(E_0 - E_e)}{E_e} - \frac{1}{24}\,\frac{(E_0 -
  E_e)^2}{ E^2_e}\Big)\nonumber\\
\hspace{-0.3in}&& \times\,\Big[\frac{1}{\beta}\,{\ell n}\Big(\frac{1 +
    \beta}{1 - \beta}\Big) - 2\Big] - \frac{1}{12}\,\frac{(E_0 - E_e)^2}{ E^2_e}\Big\} -
\Big\{2\,E_e (E_0 - E_e)^2-
\frac{1}{6}\,\frac{(E_0 - E_e)^5 + 2 (k^2_p + k^2_e)(E_0 -
  E_e)^3}{\beta^2 E^2_e}\nonumber\\
\hspace{-0.3in}&&- (k^2_p - k^2_e)\,(E_0 - E_e)
\Big\}\,\Big[\frac{1}{\beta}\,{\ell n}\Big(\frac{1 + \beta}{1 -
    \beta}\Big) - 2\Big] - \frac{1}{12}\,\frac{(E_0 - E_e)^4 - 3 (k^2_p -
  k^2_e)\,(E_0 - E_e)^2}{E_e}\,\frac{1}{\beta}\,{\ell n}\Big(\frac{1 +
  \beta}{1 - \beta}\Big)\nonumber\\
\hspace{-0.3in}&& + \Big\{\frac{1}{6}\,\frac{((E_0 - E_e)^4 + 3 (k^2_p
  + k^2_e)(E_0 - E_e)^2}{\beta^2 E_e} - \frac{(k^2_p + k^2_e)(E_0 -
  E_e)}{\beta^2 }\Big\}\,\Big[\frac{3 - \beta^2}{\beta}\,{\ell
    n}\Big(\frac{1 + \beta}{1 - \beta}\Big) - 6\Big]\Bigg\}\nonumber\\
\hspace{-0.3in}&&\times\,F(E_e, Z = 1) = - 0.020624\,{\rm
  MeV^5}.
\end{eqnarray}
For the calculation of the contributions of the soft--photons we have
followed the paper by Gl\"uck \cite{Gluck1997} and restricted the
photon--energy spectrum from above at $\omega_m$. For numerical
calculations we have set $\omega_m = (E_0 - m_e)/3$. The limits of the
integration, plotted in Fig.\,1, are equal to
\begin{eqnarray}\label{label21}
\hspace{-0.3in} k_1 &=& (E_m - E_0) + \sqrt{E^2_m -
  m^2_e}\;\;,\;\; k_2 = (E_0 - E_m) + \sqrt{E^2_m -
  m^2_e} \;\;,\;\;  (k_p)_{\rm max} = \sqrt{E^2_0 - m^2_e},\nonumber\\
\hspace{-0.3in} (E_e)_{\rm min} &=& \frac{(E_0 - k_p)^2 + m^2_e}{2
  (E_0 - k_p)} \;\;,\;\; (E_e)_{\rm max} = \frac{(E_0 + k_p)^2 +
  m^2_e}{2 (E_0 + k_p)}\;,\; E_0 =  \frac{m^2_n - m^2_p + m^2_e}{2
  m_n} \;\;,\;\; E_m = E_0 - \omega_m.
\end{eqnarray} 
\begin{figure}
\centering \includegraphics[height=0.35\textheight]{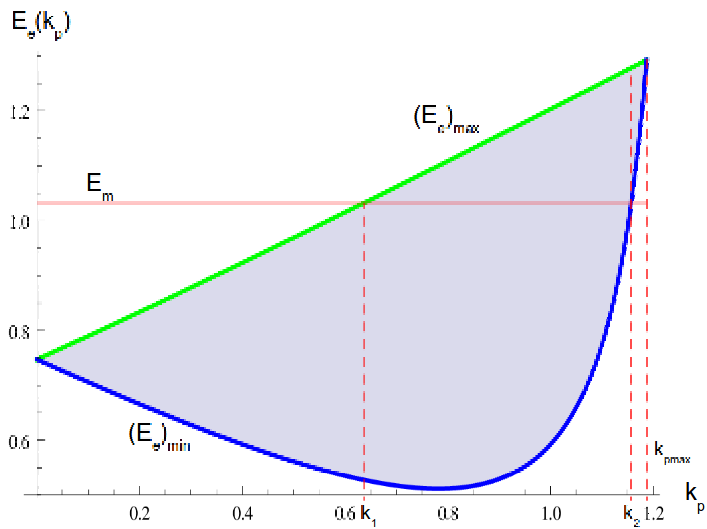}
\caption{The energy regions of the integration over the electron
  energy $E_e$ and the proton momentum $k_p$. The regions above and
  below $E_m = E_0 - \omega_m$ correspond to the soft- and hard-photon
  energy region, respectively.}
\end{figure}
The shaded regions above and below $E_m = E_0 - \omega_m$ correspond
to the soft- and hard-photon energy region, respectively. The function
$(E_e)_{\rm max}$ is practically a straight line. It is not a surprise,
since the function $(E_e)_{\rm max}$ can be written as
\begin{figure}
\centering \includegraphics[height=0.30\textheight]{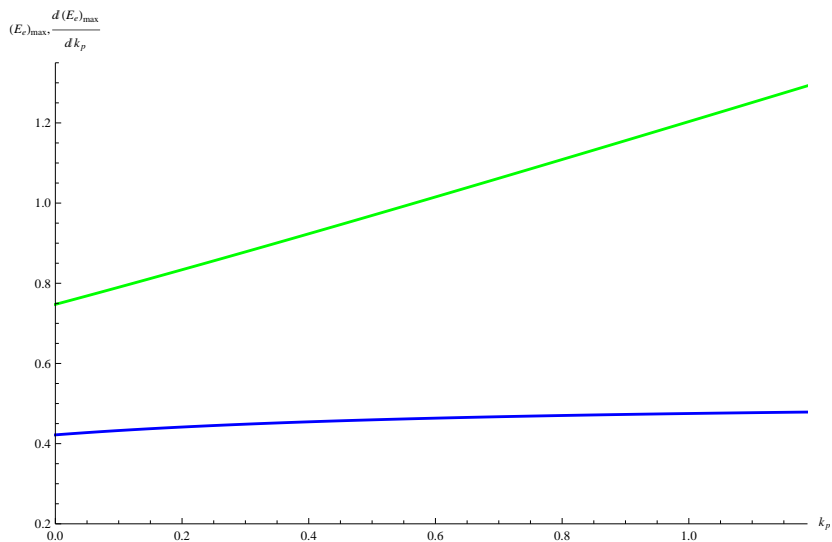}
\caption{The maximal electron energy $(E_e)_{\rm max}$ (green line)
  and its derivative (blue line) as functions of $k_p$ for $0 \le k_p
  \le (k_p)_{\rm max}$.}
\end{figure}
\begin{eqnarray*}\label{label21a}
\hspace{-0.3in} (E_e)_{\rm max} = \frac{(E_0 + k_p)^2 + m^2_e}{2 (E_0
  + k_p)} = \frac{1}{2}\Big(E_0 + \frac{m^2_e}{E_0 + k_p}\Big) + \frac{1}{2}\,k_p,
\end{eqnarray*} 
where $E_0 \gg m^2_e/(E_0 + k_p)$ for $0 \le k_p \le (k_p)_{\rm max} =
\sqrt{E^2_0 - m^2_e}$. In Fig.\,2 we plot $(E_e)_{\rm max}$ and the
derivative $d(E_e)_{\rm max}/dk_p$.  One may see that the derivative
of $(E_e)_{\rm max}$ is practically constant. This confirms a
behaviour of $(E_e)_{\rm max}$ as a straight line.

\section{Hard-photon contribution to proton recoil energy and angular 
distribution of neutron radiative $\beta^-$--decay}
\label{sec:hardphoton}

In this section we calculate the contributions of the hard-photons
with energies $\omega \ge \omega_m$. In this photon-energy region the
functions $g^{(1)}_{\beta^-_c\gamma}(E_e,\mu)$ and
$g^{(2)}_{\beta^-_c\gamma}(E_e,\mu)$ should be replaced by the
functions $g^{(1)}_{\beta^-_c\gamma}(E_e,\omega_m)$ and
$g^{(2)}_{\beta^-_c\gamma}(E_e, \omega_m)$, respectively, defined in
Appendix B of Ref.\cite{Ivanov2012} (see Eq.(B.15)). They are given by
\begin{eqnarray}\label{label22}
\hspace{-0.3in}&&g^{(1)}_{\beta^-_c\gamma}(E_e,\omega_m) = \Big\{{\ell
  n}\Big(\frac{E_0 - E_e}{\omega_m}\Big) - \frac{3}{2} + \frac{2
  \omega_m}{E_0 - E_e} - \frac{1}{2}\,\frac{\omega^2_m}{(E_0 - E_e)^2}
+ \frac{1}{3}\,\frac{(E_0 - E_e - \omega_m)^3}{E_e(E_0 -
  E_e)^2}\nonumber\\
\hspace{-0.3in}&& + \frac{1}{24}\,\frac{(E_0 - E_e -
  \omega_m)^3}{E^2_e(E_0 - E_e)^2}\,(E_0 - E_e + 3\omega_m)
\Big\}\,\Big[\frac{1}{\beta}\,{\ell n}\Big(\frac{1 + \beta}{1 -
    \beta}\Big) - 2\Big] + \frac{1}{12}\,\frac{(E_0 - E_e -
  \omega_m)^3}{E^2_e(E_0 - E_e)^2}\,(E_0 - E_e + 3\omega_m),\nonumber\\
\hspace{-0.3in}&&g^{(2)}_{\beta^-_c\gamma}(E_e,\omega_m) = \Big\{{\ell
  n}\Big(\frac{E_0 - E_e}{\omega_m}\Big) - \frac{3}{2} + \frac{2
  \omega_m}{E_0 - E_e} - \frac{1}{2}\,\frac{\omega^2_m}{(E_0 - E_e)^2} + 
\frac{1}{3}\,\frac{(E_0 - E_e - \omega_m)^3}{\beta^2 E_e(E_0 - E_e)^2}\nonumber\\
\hspace{-0.3in}&& + \frac{1}{24}\,\frac{(E_0 - E_e -
  \omega_m)^3}{\beta^2 E^2_e(E_0 - E_e)^2}\,(E_0 - E_e + 3\omega_m)
\Big\} \,\Big[\frac{1}{\beta}\,{\ell n}\Big(\frac{1 + \beta}{1 -
    \beta}\Big) - 2\Big].
\end{eqnarray}
In terms of the functions $g^{(1)}_{\beta^-_c\gamma}(E_e,\omega_m)$
and $g^{(2)}_{\beta^-_c\gamma}(E_e, \omega_m)$ we transcribe
Eq.(\ref{label6}) into the form
\begin{eqnarray}\label{label23}
\hspace{-0.3in}&&\frac{d^3\Delta \lambda^{(1)}_{\beta^-_c}(E_e, k_p,
  \theta_p, P)}{dE_e dk_pd\cos\theta_p} = (1 +
3\lambda^2)\,\frac{\alpha}{\pi}\,\frac{G^2_F|V_{ud}|^2}{8\pi^3}\,\Big\{-\,
\Big((E_0 - E_e) - P B_0\,\cos\theta_p\,\frac{(E_0 - E_e)^2 + k^2_p - k^2_e}{2 k_p
  }\Big)\,g^{(1)}_{\beta^-_c\gamma}(E_e,\omega_m)\nonumber\\
\hspace{-0.3in}&& - \,\Big(- a_0\,\frac{(E_0 - E_e)^2 - k^2_p +
  k^2_e}{2 E_e} + P A_0\,\cos\theta_p\,(E_0 - E_e)\,\frac{(E_0 -
  E_e)^2 - k^2_p - k^2_e}{2 k_p
  E_e}\Big)\,g^{(2)}_{\beta^-_c\gamma}(E_e,\omega_m)\Big\}\,E_e
k_p\,F(E_e, Z = 1).\nonumber\\
\hspace{-0.3in}&&
\end{eqnarray}
In the hard--photon energy region we analyse the sum of
Eqs.(\ref{label3}) and (\ref{label4}). The result is
\begin{eqnarray}\label{label24}
\hspace{-0.3in}&&\sum_{j = 2,3}\frac{d^3\Delta \lambda^{(j)}_{\beta^-_c}(E_e, k_p,
  \theta_p, P)}{dE_e dk_pd\cos\theta_p} = (1 +
3\lambda^2)\,\frac{\alpha}{\pi}\,\frac{G^2_F|V_{ud}|^2}{4\pi^3}
\int^{2\pi}_0 \frac{d\phi_p}{2\pi} \int \frac{d\Omega_{ep}}{4\pi} \int
\frac{d\omega}{\omega}\int \frac{d\Omega_{\gamma}}{4\pi}\,\Big\{\Big(1
- B_0\,\frac{\vec{\xi}_n \cdot (\vec{k}_p + \vec{k}_e +
  \omega\,\vec{n}\,)}{|\vec{k}_p + \vec{k}_e +
  \omega\,\vec{n}\,|}\Big)\nonumber\\
\hspace{-0.3in}&&\times\,\Big[\frac{k^2_e - (\vec{n}\cdot
  \vec{k}_e)^2}{(E_e - \vec{n}\cdot \vec{k}_e)^2}\,\Big(1 +
\frac{\omega}{E_e}\Big) + \frac{1}{E_e - \vec{n}\cdot
  \vec{k}_e}\,\frac{\omega^2}{E_e}\Big] + \Big(- a_0\,\frac{ \vec{k}_p + \vec{k}_e
  + \omega\,\vec{n}}{|\vec{k}_p + \vec{k}_e +
  \omega\,\vec{n}\,|} + A_0\,\vec{\xi}_n\Big)\cdot
\Big[\Big(\frac{k^2_e - (\vec{n}\cdot \vec{k}_e)^2}{(E_e -
    \vec{n}\cdot \vec{k}_e)^2} + \frac{\omega}{E_e -\vec{n}\cdot
    \vec{k}_e }\Big)\nonumber\\
\hspace{-0.3in}&&\times\,\frac{\vec{k}_e}{E_e} + \Big( -
\frac{m^2_e}{(E_e - \vec{n}\cdot \vec{k}_e)^2} + \frac{E_e +
  \omega}{E_e - \vec{n}\cdot
  \vec{k}_e}\Big)\,\frac{\omega}{E_e}\,\vec{n}\Big]\Big\}\,\delta(E_0
- E_e - |\vec{k}_p + \vec{k}_e + \omega\,\vec{n}| - \omega)\,k_e E_e
k^2_p\,F(E_e, Z = 1).
\end{eqnarray}
Using energy conservation we obtain
\begin{eqnarray}\label{label25}
\hspace{-0.3in}&&\sum_{j = 2,3}\frac{d^3\Delta \lambda^{(j)}_{\beta^-_c}(E_e, k_p,
  \theta_p, P)}{dE_e dk_pd\cos\theta_p} = (1 +
3\lambda^2)\,\frac{\alpha}{\pi}\,\frac{G^2_F|V_{ud}|^2}{2\pi^3}
\int^{2\pi}_0 \frac{d\phi_p}{2\pi} \int \frac{d\Omega_{ep}}{4\pi} \int
\frac{d\omega}{\omega}\int \frac{d\Omega_{\gamma}}{4\pi}\,\Big\{
\Big((E_0 - E_e - \omega)\nonumber\\
\hspace{-0.3in}&& - B_0\,\vec{\xi}_n \cdot (\vec{k}_p + \vec{k}_e +
\omega\,\vec{n}\,)\Big)\,\Big[\frac{k^2_e - (\vec{n}\cdot
    \vec{k}_e)^2}{(E_e - \vec{n}\cdot \vec{k}_e)^2}\,\Big(1 +
  \frac{\omega}{E_e}\Big) + \frac{1}{E_e - \vec{n}\cdot
    \vec{k}_e}\,\frac{\omega^2}{E_e}\Big] + \Big(- a_0\,( \vec{k}_p +
\vec{k}_e + \omega\,\vec{n})\nonumber\\
\hspace{-0.3in}&& + A_0\,(E_0 - E_e - \omega)\,\vec{\xi}_n\Big)\cdot
\Big[\Big(\frac{k^2_e - (\vec{n}\cdot \vec{k}_e)^2}{(E_e -
    \vec{n}\cdot \vec{k}_e)^2} + \frac{\omega}{E_e -\vec{n}\cdot
    \vec{k}_e }\Big)\,\frac{\vec{k}_e}{E_e} + \Big( - \frac{m^2_e}{(E_e - \vec{n}\cdot
  \vec{k}_e)^2} + \frac{E_e + \omega}{E_e - \vec{n}\cdot
  \vec{k}_e}\Big)\,\frac{\omega}{E_e}\,\vec{n}\Big]\Big\}\nonumber\\
\hspace{-0.3in}&& \times\,\delta\Big((E_0 - E_e - \omega)^2 -
(\vec{k}_p + \vec{k}_e + \omega\,\vec{n})^2\Big)\,k_e E_e
k^2_p\,F(E_e, Z = 1).
\end{eqnarray}
Since an analytical calculation of the integrals in
Eq.(\ref{label25}) is not practically possible, we proceed to a
numerical calculation. For this aim we define the scalar products in
terms of the angular variables. We set
\begin{eqnarray*}
\hspace{-0.3in}\vec{\xi}_n\cdot \vec{k}_p &=& P k_p
\,\cos\theta_p,\nonumber\\
\hspace{-0.3in}\vec{\xi}_n\cdot \vec{k}_e &=& P k_e
\,(\cos\theta_p \cos\theta_{ep} +
\sin\theta_p\sin\theta_{ep}\,\cos(\phi_p -
\phi_{ep})),\nonumber\\ 
\hspace{-0.3in}\vec{\xi}_n\cdot \vec{n} &=& P\,(\cos\theta_p \cos\theta_{p\gamma} +
\sin\theta_p\sin\theta_{p\gamma}\,\cos(\phi_p -
\phi_{p\gamma})),
\end{eqnarray*}
\begin{eqnarray}\label{label26}
\hspace{-0.3in}\vec{k}_p \cdot \vec{k}_e &=& k_p k_e\,\cos\theta_{ep},\nonumber\\
\hspace{-0.3in}\vec{k}_p\cdot \vec{n} &=& k_p\,\cos\theta_{p\gamma},\nonumber\\
\hspace{-0.3in}\vec{k}_e\cdot \vec{n} &=& k_e\,(\cos\theta_{ep} \cos\theta_{p\gamma} + \sin\theta_{ep}\sin\theta_{p\gamma}\,\cos(\phi_{ep} -
\phi_{p\gamma})).
\end{eqnarray}
Since azimuthal angles enter in the differences, the azimuthal angle
$\phi_{p\gamma}$ may be excluded. This gives
\begin{eqnarray}\label{label27}
\hspace{-0.3in}\vec{\xi}_n\cdot \vec{k}_p &=& P k_p
\,\cos\theta_p,\nonumber\\ 
\hspace{-0.3in}\vec{\xi}_n\cdot \vec{k}_e &=& P k_e
\,(\cos\theta_p \cos\theta_{ep} +
\sin\theta_p\sin\theta_{ep}\,\cos(\phi_p -
\phi_{ep})),\nonumber\\ 
\hspace{-0.3in}\vec{\xi}_n\cdot \vec{n} &=& P\,(\cos\theta_p \cos\theta_{p\gamma} +
\sin\theta_p\sin\theta_{p\gamma}\,\cos\phi_p ),\nonumber\\
\hspace{-0.3in}\vec{k}_p \cdot \vec{k}_e &=& k_p k_e\,\cos\theta_{ep},\nonumber\\
\hspace{-0.3in}\vec{k}_p\cdot \vec{n} &=& k_p\,\cos\theta_{p\gamma},\nonumber\\
\hspace{-0.3in}\vec{k}_e\cdot \vec{n} &=& k_e\,(\cos\theta_{ep}
\cos\theta_{p\gamma} + \sin\theta_{ep}\sin\theta_{p\gamma}\,\cos
\phi_{ep}).
\end{eqnarray}
After the integration of Eq.(\ref{label25}) over $\omega$ in the
limits $\omega_m \le \omega \le (E_0 - m_e)/2$ we arrive at the expression
\begin{eqnarray}\label{label28}
\hspace{-0.3in}&&\sum_{j = 2,3}\frac{d^3\Delta
  \lambda^{(j)}_{\beta^-_c}(E_e, k_p, \theta_p, P)}{dE_e
  dk_pd\cos\theta_p} = (1 +
3\lambda^2)\,\frac{\alpha}{\pi}\,\frac{G^2_F|V_{ud}|^2}{16\pi^3}
\int^{2\pi}_0 \frac{d\phi_p}{2\pi} \int^{2\pi}_0
\frac{d\phi_{ep}}{2\pi} \int^{+1}_{-1}d\cos\theta_{ep}
\int^{+1}_{-1}d\cos\theta_{p\gamma}\nonumber\\
\hspace{-0.3in}&&\times\,\Bigg\{\Theta\Bigg((E_0 - m_e) - \frac{ (E_0
  - E_e)^2 - k^2_p - k^2_e - 2\vec{k}_e\cdot \vec{k}_p}{(E_0 - E_e) +
  \vec{k}_p\cdot \vec{n} + \vec{k}_e\cdot \vec{n}}\Bigg) -
\Theta\Bigg(\omega_m - \frac{1}{2}\,\frac{(E_0 - E_e)^2 - k^2_p -
  k^2_e - 2\vec{k}_e\cdot \vec{k}_p}{(E_0 - E_e) + \vec{k}_p\cdot
  \vec{n} + \vec{k}_e\cdot \vec{n}}\Bigg)\Bigg\}\nonumber\\
\hspace{-0.3in}&&\times\,\frac{1}{(E_0 - E_e) + \vec{k}_p\cdot \vec{n}
  + \vec{k}_e\cdot \vec{n}}\,\Big\{ \Big((E_0 - E_e - \omega) -
B_0\,\vec{\xi}_n \cdot (\vec{k}_p + \vec{k}_e +
\omega\,\vec{n}\,)\Big)\,\Big[\frac{k^2_e - (\vec{n}\cdot
    \vec{k}_e)^2}{(E_e - \vec{n}\cdot \vec{k}_e)^2}\,\Big(\frac{1}{\omega} +
  \frac{1}{E_e}\Big)\nonumber\\
\hspace{-0.3in}&& + \frac{1}{E_e - \vec{n}\cdot
  \vec{k}_e}\,\frac{\omega}{E_e}\Big] + \Big(- a_0\,( \vec{k}_p +
\vec{k}_e + \omega\,\vec{n}) + A_0\,(E_0 - E_e -
\omega)\,\vec{\xi}_n\Big)\cdot \Big[\Big(\frac{1}{\omega}\,\frac{k^2_e
    - (\vec{n}\cdot \vec{k}_e)^2}{(E_e - \vec{n}\cdot \vec{k}_e)^2} +
  \frac{1}{E_e -\vec{n}\cdot \vec{k}_e
  }\Big)\,\frac{\vec{k}_e}{E_e}\nonumber\\
\hspace{-0.3in}&& + \Big( - \frac{m^2_e}{(E_e - \vec{n}\cdot
  \vec{k}_e)^2} + \frac{E_e + \omega}{E_e - \vec{n}\cdot
  \vec{k}_e}\Big)\,\frac{1}{E_e}\,\vec{n}\Big]\Big\}\,k_e E_e k^2_p\,F(E_e, Z = 1),
\end{eqnarray}
where $\omega$ is the function, given by 
\begin{eqnarray}\label{label29}
\hspace{-0.3in}\omega = \frac{1}{2}\,\frac{(E_0 -
  E_e)^2 - k^2_p - k^2_e - 2\vec{k}_e\cdot \vec{k}_p}{(E_0 -
  E_e) + \vec{k}_p\cdot \vec{n} + \vec{k}_e\cdot \vec{n}\,},
\end{eqnarray} 
and $\Theta(z)$ is the Heaviside stepfunction.

Now we may define the contributions of the hard--photons to the proton
recoil angular distribution of the neutron $\beta^-$--decay.  Having
integrated Eq.(\ref{label23}) and Eq.(\ref{label28}) over the electron
energy $E_e$ and the proton momentum $k_p$ we obtain $\Delta
X^{(h)}_2$, $\Delta X^{(h)}_0$, $\Delta X^{(h)}_{10}$ and $\Delta
X^{(h)}_{11}$ equal to the following: \\ i) $\Delta X^{(h)}_2$:
\begin{eqnarray}\label{label30}
\hspace{-0.3in}\Delta X^{(h)}_2 &=& \int^{k_1}_0dk_p\int^{(E_e)_{\rm
    max}}_{(E_e)_·{\rm min}}dE_e\,\Big\{- 2\,(E_0 -
E_e)\,g^{(1)}_{\beta^-_c\gamma}(E_e,\omega_m) + \Delta
g_n(E_e, k_p)\, k_e k_p\Big\}\,E_e k_p\,F(E_e, Z = 1)\nonumber\\
\hspace{-0.3in} &+& \int^{k_2}_{k_1}dk_p\int^{E_m}_{(E_e)_·{\rm
    min}}dE_e\,\Big\{- 2\,(E_0 -
E_e)\,g^{(1)}_{\beta^-_c\gamma}(E_e,\omega_m) + \Delta g_n(E_e, k_p)\,
k_e k_p\Big\}\,E_e k_p\,F(E_e, Z = 1) =\nonumber\\
\hspace{-0.3in}&=& - 0.005268\,{\rm MeV^5}
\end{eqnarray}
with 
\begin{eqnarray}\label{label31}
\hspace{-0.3in}&&\Delta g_n(E_e, k_p) = \int^{2\pi}_0
\frac{d\phi_{ep}}{2\pi} \int^{+1}_{-1}d\cos\theta_{ep}
\int^{+1}_{-1}d\cos\theta_{p\gamma}\,\Bigg\{\Theta\Bigg((E_0 - m_e) -
\frac{(E_0 - E_e)^2 - k^2_p - k^2_e - 2\vec{k}_e\cdot \vec{k}_p}{(E_0
  - E_e) + \vec{k}_p\cdot \vec{n} + \vec{k}_e\cdot \vec{n}}\Bigg)\nonumber\\
\hspace{-0.3in}&& - \Theta\Bigg(\omega_m - \frac{1}{2}\,\frac{(E_0 -
  E_e)^2 - k^2_p - k^2_e - 2\vec{k}_e\cdot \vec{k}_p}{(E_0 - E_e) +
  \vec{k}_p\cdot \vec{n} + \vec{k}_e\cdot
  \vec{n}}\Bigg)\Bigg\}\,\frac{(E_0 - E_e - \omega)}{(E_0 - E_e) +
  \vec{k}_p\cdot \vec{n} + \vec{k}_e\cdot \vec{n}}\,\Bigg\{\frac{k^2_e - (\vec{n}\cdot
  \vec{k}_e)^2}{(E_e - \vec{n}\cdot
  \vec{k}_e)^2}\,\Big(\frac{1}{\omega} + \frac{1}{E_e}\Big)\nonumber\\
\hspace{-0.3in}&& + \frac{1}{E_e - \vec{n}\cdot
  \vec{k}_e}\,\frac{\omega}{E_e}\Bigg\},
\end{eqnarray}
ii) $\Delta X^{(h)}_0$:
\begin{eqnarray}\label{label32}
\hspace{-0.3in}\Delta X^{(h)}_0 &=& \int^{k_1}_0dk_p\int^{(E_e)_{\rm
    max}}_{(E_e)_·{\rm min}}dE_e\,\Big\{\frac{(E_0 - E_e)^2 - k^2_p +
  k^2_e}{E_e}\,g^{(2)}_{\beta^-_c\gamma}(E_e,\omega_m) + \Delta
f_a (E_e, k_p)\, k_e k_p\Big\}\,E_e k_p\,F(E_e, Z = 1)\nonumber\\
\hspace{-0.3in} &+& \int^{k_2}_{k_1}dk_p\int^{E_m}_{(E_e)_·{\rm
    min}}dE_e\,\Big\{\frac{(E_0 - E_e)^2 - k^2_p +
  k^2_e}{E_e}\,g^{(2)}_{\beta^-_c\gamma}(E_e,\omega_m) + \Delta
f_a(E_e, k_p)\,  k_e k_p\Big\}\,E_e k_p\,F(E_e, Z = 1) =\nonumber\\
\hspace{-0.3in}&=& 0.000049\,{\rm MeV^5}
\end{eqnarray}
with 
\begin{eqnarray}\label{label33}
\hspace{-0.3in}&&\Delta f_a(E_e, k_p) = \int^{2\pi}_0
\frac{d\phi_{ep}}{2\pi} \int^{+1}_{-1}d\cos\theta_{ep}
\int^{+1}_{-1}d\cos\theta_{p\gamma}\,\Bigg\{\Theta\Bigg((E_0 - m_e) -
\frac{(E_0 - E_e)^2 - k^2_p - k^2_e - 2\vec{k}_e\cdot \vec{k}_p}{(E_0
  - E_e) + \vec{k}_p\cdot \vec{n} + \vec{k}_e\cdot \vec{n}}\Bigg)\nonumber\\
\hspace{-0.3in}&& - \Theta\Bigg(\omega_m - \frac{1}{2}\,\frac{(E_0 -
  E_e)^2 - k^2_p - k^2_e - 2\vec{k}_e\cdot \vec{k}_p}{(E_0 - E_e) +
  \vec{k}_p\cdot \vec{n} + \vec{k}_e\cdot
  \vec{n}}\Bigg)\Bigg\}\,\frac{1}{(E_0 - E_e) + \vec{k}_p\cdot \vec{n}
  + \vec{k}_e\cdot \vec{n}}\,\Bigg\{ - \frac{\vec{k}_p\cdot \vec{k}_e +
  k^2_e + \omega \vec{n}\cdot
  \vec{k}_e}{E_e}\nonumber\\
\hspace{-0.3in}&& \times\,\Bigg(\frac{1}{\omega}\,\frac{k^2_e -
  (\vec{n}\cdot \vec{k}_e)^2}{(E_e - \vec{n}\cdot \vec{k}_e)^2} +
\frac{1}{E_e -\vec{n}\cdot \vec{k}_e }\Bigg) - \frac{\vec{k}_p\cdot
  \vec{n} + \vec{k}_e\cdot \vec{n} + \omega}{E_e}\Bigg( -
\frac{m^2_e}{(E_e - \vec{n}\cdot \vec{k}_e)^2}+ \frac{E_e +
  \omega}{E_e - \vec{n}\cdot \vec{k}_e}\Bigg)\Bigg\},
\end{eqnarray}
iii) $\Delta X^{(h)}_{10}$:
\begin{eqnarray}\label{label34}
\hspace{-0.3in}\Delta X^{(h)}_{10} &=& \int^{k_1}_0dk_p\int^{(E_e)_{\rm
    max}}_{(E_e)_·{\rm min}}dE_e\,\Big\{(E_0 - E_e)\,\frac{(E_0 -
  E_e)^2 - k^2_p - k^2_e}{k_p
  E_e}\,g^{(2)}_{\beta^-_c\gamma}(E_e,\omega_m) + \Delta
f_A(E_e, k_p)\,  k_e k_p\Big\}\nonumber\\
\hspace{-0.3in} &&\times\,E_e k_p\,F(E_e, Z = 1)\nonumber\\
\hspace{-0.3in}&+&\int^{k_2}_{k_1}dk_p\int^{E_m}_{(E_e)_·{\rm
    min}}dE_e\,\Big\{(E_0 - E_e)\,\frac{(E_0 -
  E_e)^2 - k^2_p - k^2_e}{k_p
  E_e}\,g^{(2)}_{\beta^-_c\gamma}(E_e,\omega_m) + \Delta
f_A (E_e, k_p)\, k_e k_p\Big\}\nonumber\\
\hspace{-0.3in} &&\times\,E_e k_p\,F(E_e, Z = 1) = - 0.012227\,{\rm
  MeV^5}
\end{eqnarray}
with 
\begin{eqnarray}\label{label35}
\hspace{-0.3in}&&\Delta f_A(E_e, k_p) = \int^{2\pi}_0
\frac{d\phi_{ep}}{2\pi} \int^{+1}_{-1}d\cos\theta_{ep}
\int^{+1}_{-1}d\cos\theta_{p\gamma}\,\Bigg\{\Theta\Bigg((E_0 - m_e) -
\frac{(E_0 - E_e)^2 - k^2_p - k^2_e - 2\vec{k}_e\cdot \vec{k}_p}{(E_0
  - E_e) + \vec{k}_p\cdot \vec{n} + \vec{k}_e\cdot \vec{n}}\Bigg)\nonumber\\
\hspace{-0.3in}&& - \Theta\Bigg(\omega_m - \frac{1}{2}\,\frac{(E_0 -
  E_e)^2 - k^2_p - k^2_e - 2\vec{k}_e\cdot \vec{k}_p}{(E_0 - E_e) +
  \vec{k}_p\cdot \vec{n} + \vec{k}_e\cdot
  \vec{n}}\Bigg)\Bigg\}\,\frac{(E_0 - E_e - \omega)}{(E_0 - E_e) +
  \vec{k}_p\cdot \vec{n} + \vec{k}_e\cdot \vec{n}}\nonumber\\
\hspace{-0.3in}&&\times\,\Bigg\{\cos\theta_{ep}\,\frac{k_e}{E_e}\Bigg(\frac{1}{\omega}\,\frac{k^2_e -
  (\vec{n}\cdot \vec{k}_e)^2}{(E_e - \vec{n}\cdot \vec{k}_e)^2} +
\frac{1}{E_e -\vec{n}\cdot \vec{k}_e }\Bigg) +
\cos\theta_{p\gamma}\,\frac{1}{E_e}\Bigg( - \frac{m^2_e}{(E_e -
  \vec{n}\cdot \vec{k}_e)^2} + \frac{E_e + \omega}{E_e - \vec{n}\cdot
  \vec{k}_e}\Bigg)\Bigg\}
\end{eqnarray}
and  iv) $\Delta X^{(h)}_{11}$:
\begin{eqnarray}\label{label36}
\hspace{-0.3in}\Delta X^{(h)}_{11} &=&
\int^{k_1}_0dk_p\int^{(E_e)_{\rm max}}_{(E_e)_·{\rm min}}dE_e\,\Big\{-
\frac{(E_0 - E_e)^2 + k^2_p - k^2_e}{k_p
}\,g^{(1)}_{\beta^-_c\gamma}(E_e,\omega_m) + \Delta f_B(E_e, k_p)\,
k_e k_p\Big\}\,E_e k_p\,F(E_e, Z = 1)\nonumber\\
\hspace{-0.3in} &+& \int^{k_2}_{k_1}dk_p\int^{E_m}_{(E_e)_·{\rm
    min}}dE_e\,\Big\{- \frac{(E_0 - E_e)^2 + k^2_p - k^2_e}{k_p
}\,g^{(1)}_{\beta^-_c\gamma}(E_e,\omega_m) + \Delta f_B(E_e, k_p)\,  k_e
k_p\Big\}\,E_e k_p\,F(E_e, Z = 1) =\nonumber\\
\hspace{-0.3in}&=& - 0.007939\,{\rm MeV^5}
\end{eqnarray}
with
\begin{eqnarray}\label{label37}
\hspace{-0.3in}&&\Delta f_B(E_e, k_p) = \int^{2\pi}_0
\frac{d\phi_{ep}}{2\pi} \int^{+1}_{-1}d\cos\theta_{ep}
\int^{+1}_{-1}d\cos\theta_{p\gamma}\,\Bigg\{\Theta\Bigg((E_0 - m_e) -
\frac{(E_0 - E_e)^2 - k^2_p - k^2_e - 2\vec{k}_e\cdot \vec{k}_p}{(E_0
  - E_e) + \vec{k}_p\cdot \vec{n} + \vec{k}_e\cdot \vec{n}}\Bigg)\nonumber\\
\hspace{-0.3in}&& - \Theta\Bigg(\omega_m - \frac{1}{2}\,\frac{(E_0 -
  E_e)^2 - k^2_p - k^2_e - 2\vec{k}_e\cdot \vec{k}_p}{(E_0 - E_e) +
  \vec{k}_p\cdot \vec{n} + \vec{k}_e\cdot
  \vec{n}}\Bigg)\Bigg\}\,\frac{1}{(E_0 - E_e) + \vec{k}_p\cdot \vec{n}
  + \vec{k}_e\cdot \vec{n}}\nonumber\\
\hspace{-0.3in}&&\times\,\Big(k_p + k_e \cos\theta_{ep} +
\omega\,\cos\theta_{p\gamma}\Big)\Bigg\{\frac{k^2_e - (\vec{n}\cdot
  \vec{k}_e)^2}{(E_e - \vec{n}\cdot
  \vec{k}_e)^2}\,\Big(\frac{1}{\omega} + \frac{1}{E_e}\Big) +
\frac{1}{E_e - \vec{n}\cdot \vec{k}_e}\,\frac{\omega}{E_e}\Bigg\}.
\end{eqnarray}
Now let us summarise the obtained results.

\section{Conclusion}
\label{sec:conclusion}

We have analysed the contributions of the proton--photon correlations
to the proton recoil energy and angular distribution of the neutron
$\beta^-$--decay, calculated in \cite{Ivanov2012}. As has been shown
in \cite{Ivanov2012}, the radiative corrections to the proton recoil
energy and angular distribution can be described by the functions
$(\alpha/\pi)\,g_n(E_e)$ and $(\alpha/\pi)\,f_n(E_e)$, which were
calculated by neglecting the proton--photon correlations in the
radiative $\beta^-$--decay of the neutron. As has been pointed out by
Gl\"uck \cite{Gluck1997}, the problem of the proton--photon
correlations in the proton recoil energy spectrum of the radiative
$\beta$--decay of the neutron and the contributions of these
correlations to the energy and angular distributions of the neutron
$\beta^-$--decay should be thoroughly investigated by means of a
numerical analysis of the hard-photon energy region. For the
calculation of the contributions of the hard photons Gl\"uck used the
Monte Carlo simulation method \cite{Gluck1997}. For the analysis of
the contributions of the proton--photon correlations in the radiative
$\beta^-$--decay of the neutron we have defined a correction to the
proton recoil energy and angular distribution of the neutron
$\beta^-$--decay from the proton--photon correlations in the radiative
$\beta^-$--decay of the neutron.  In our analysis of this correction
we have followed the paper by Gl\"uck \cite{Gluck1997}. We have
divided the photon--energy spectrum into two parts, corresponding to
the soft and hard photons. The contribution of the soft photons we
have calculated analytically, whereas the contribution of the hard
photons has been calculated numerically. For this aim we have used
MATHEMATICA by Wolfram \cite{Wolfram}. The soft- and hard-photon
energy regions we have divided by $\omega_m$. For numerical
calculations we have set $\omega_m = (E_0 - m_e)/3 = 0.260\,{\rm
  MeV}$. Integrating over the electron energy $E_e$ and the proton
momentum $k_p$ we have obtained the contributions, caused by the
proton--photon correlations in the radiative $\beta^-$--decay of the
neutron, to the proton angular distribution of the neutron
$\beta^-$--decay
\begin{eqnarray}\label{label38}
\hspace{-0.3in}&&\frac{d\Delta \lambda_{\beta^-_c}( \theta_p,
  P)}{d\cos\theta_p} = (1 +
3\lambda^2)\,\frac{G^2_F|V_{ud}|^2}{16\pi^3}
\,\Big\{\frac{\alpha}{\pi}\,\Delta X^{(s + h)}_2 +
a_0\,\frac{\alpha}{\pi}\,\Delta X^{(s + h)}_0 + P
\cos\theta_p\,\Big(A_0\,\frac{\alpha}{\pi}\,\Delta X^{(s + h)}_{10} -
B_0\,\frac{\alpha}{\pi}\,\Delta X^{(s + h)}_{11}\Big)\Big\},\nonumber\\
\hspace{-0.3in}&&
\end{eqnarray}
where $\Delta X^{(s + h)}_2$, $\Delta X^{(s + h)}_0 $, $\Delta X^{(s +
  h)}_{10}$ and $\Delta X^{(s + h)}_{11}$ , defined by the
contributions of the soft and hard photons, are equal to
\begin{eqnarray}\label{label39}
\hspace{-0.3in}\Delta X^{(s + h)}_2 &=& + 0.010783\,{\rm MeV^5},\nonumber\\
\hspace{-0.3in}\Delta X^{(s + h)}_0  &=& - 0.010152\,{\rm MeV^5},\nonumber\\
\hspace{-0.3in}\Delta X^{(s + h)}_{10} &=&  - 0.057472\,{\rm MeV^5},\nonumber\\
\hspace{-0.3in}\Delta X^{(s + h)}_{11} &=& - 0.028563\,{\rm MeV^5}.
\end{eqnarray}
The contributions of $\Delta X^{(s + h)}_2$ and $\Delta X^{(s + h)}_0
$, multiplied by $\alpha/\pi$, to the proton recoil energy and angular
distribution are of order $2.4\times 10^{-5}$ and can be neglected at
the level of accuracy $10^{-5}$, accepted in \cite{Ivanov2012}. Such a
neglect of the contributions of $\Delta X^{(s + h)}_2$ and $\Delta
X^{(s + h)}_0 $ confirms also the correctness of the use of the
radiative corrections, described by the functions
$(\alpha/\pi)\,g_n(E_e)$ and $(\alpha/\pi)\,f_n(E_e)$, for the
analysis of the proton--energy spectrum $a(T_p)$, pointed out in
\cite{Ivanov2012}. The contributions of $\Delta X^{(s + h)}_{10}$ and
$\Delta X^{(s + h)}_{11}$ to the proton recoil asymmetry $C$,
multiplied by $\alpha/\pi$, are equal to $(\alpha/\pi)\,\Delta X^{(s +
  h)}_{10} = - 1.335\times 10^{-4}$ and $(\alpha/\pi)\,\Delta X^{(s +
  h)}_{11} = - 0.664 \times 10^{-4}$. Since these corrections are of
order $10^{-4}$ and larger than accuracy of about $10^{-5}$
\cite{Ivanov2012}, they should be taken into account. This means that
in the proton recoil angular distribution of the neutron
$\beta^-$--decay, defined by Eq.(I-21) in Ref.\cite{Ivanov2012}, the
parameters $X_{10}$ and $X_{11}$ should be replaced by
\begin{eqnarray}\label{label40}
\hspace{-0.3in}&&X_{10} \to \bar{X}_{10} =  X_{10} + \Delta X^{(s + h)}_{10} = -
2.214586\,{\rm MeV^5},\nonumber\\
\hspace{-0.3in}&&X_{11} \to \bar{X}_{11} =  X_{11} + \Delta X^{(s + h)}_{11} = + 
2.215636\,{\rm MeV^5},
\end{eqnarray}
where $X_{10} = - 2.157114\,{\rm MeV^5}$ and $X_{11} = 2.244201\,{\rm
  MeV^5}$ \cite{Ivanov2012}. Denoting the new parameters as
$\bar{X}_{10} = - X_{\rm eff}$ and $\bar{X}_{11} = X_{\rm eff}$, where
$X_{\rm eff} = 2.215111\,{\rm MeV^5}$ is valid with an accuracy of
about $5.5\times 10^{-5}\,\%$ in the proton recoil angular
distribution, we may transcribe the proton recoil angular distribution
of the neutron $\beta^-$--decay, calculated in \cite{Ivanov2012}, into
the form
\begin{eqnarray}\label{label41}
\hspace{-0.3in}&&\frac{d \lambda_{\beta^-_c}(\theta_p,
  P)}{d\cos\theta_p} = (1 +
3\lambda^2)\,\frac{G^2_F|V_{ud}|^2}{16\pi^3}\,\Big\{X_1 +
\frac{\alpha}{\pi}\,X_2 +  \frac{1}{M}\,\frac{1}{1 +
  3\lambda^2}\,\Big[X_3 + (1 + 3\lambda^2)\, (X_4 + Y_1) - (1 -
\lambda^2)\, (X_5 + Y_2)\nonumber\\
\hspace{-0.3in}&& + \Big(\lambda^2 + 2(\kappa + 1)\lambda\Big)\, X_6 -
\Big(\lambda^2 - 2(\kappa + 1)\lambda\Big)\,X_7\Big] + P
\cos\theta_p\Big[ -(A_0 + B_0)\,\Big(X_8 + \frac{\alpha}{\pi}\,X_{\rm eff}\Big) + A_0\,X_9 \nonumber\\
\hspace{-0.3in}&& + \frac{1}{M}\,\frac{1}{1 + 3\lambda^2}\,\Big(
\lambda\,X_{12} - (\kappa + 1)\lambda\,X_{13} - (2\kappa +
1)\,\lambda\,X_{14}- \lambda\,(1 + \lambda)\,(X_{15} + Y_3) +
\lambda\,(1 - \lambda)\,(X_{16} + Y_4)\Big)\Big]\Big\}.
\end{eqnarray}
This leads to the following change of the correlation coefficient $C$
\begin{eqnarray}\label{label42}
\hspace{-0.3in} &&C = - \Big(x_C + \frac{\alpha}{\pi}\,x_{\rm
  eff}\Big)\,(A_0 + B_0) + \frac{1}{2}\frac{X_9}{X_1}\,A_0 +
\frac{1}{M}\,\frac{1}{1 +
  3\lambda^2}\,\Big(\lambda\,\frac{1}{2}\frac{X_{12}}{X_1} - (\kappa +
1)\,\lambda\,\frac{1}{2}\frac{X_{13}}{X_1} - (2\kappa + 1)\,\lambda\,\frac{1}{2}\frac{X_{14}}{X_1}\nonumber\\
\hspace{-0.3in}&& -\lambda(1 + \lambda)\,\frac{1}{2}\frac{X_{15} +
  Y_3}{X_1} + \lambda(1 - \lambda)\,\frac{1}{2}\frac{X_{16} +
  Y_4}{X_1}\Big) + (A_0 + B_0)\,\frac{X_8}{X_1}\,\Big\{
\frac{\alpha}{\pi}\,\frac{1}{2}\frac{X_2}{X_1} +
\frac{1}{M}\,\frac{1}{1 + 3\lambda^2}\,\Big(\frac{1}{2}\frac{X_3}{X_1}
+ (1 + 3\lambda^2)\nonumber\\
\hspace{-0.3in}&&\times\,\frac{1}{2}\frac{X_4 + Y_1}{X_1} - (1 -
\lambda^2)\, \frac{1}{2}\frac{X_5 + Y_2}{X_1}+ (\lambda^2 + 2(\kappa +
1)\lambda)\,\frac{1}{2}\frac{X_6}{X_1} - (\lambda^2 - 2(\kappa +
1)\lambda)\,\frac{1}{2}\frac{X_7}{X_1}\Big)\Big\},
\end{eqnarray}
where the contribution of the radiative corrections is symmetric with
respect to a change $A_0 \longleftrightarrow B_0$ as well as the main
term $ - x_C (A_0 + B_0)$, which has been calculated for the first
time by Treiman \cite{Treiman1958}.  The factor $x_C = X_8/2X_1 =
0.27591$, calculated in \cite{Ivanov2012}, agrees well with the factor
$x_C = 0.27594$, calculated by Gl\"uck \cite{Gluck1996}, and $x_{\rm
  eff} = X_{\rm eff}/2X_1 = 4.712120$. We would like to remind the
reader that the appearance of the term $A_0X_9/2X_1$, violating a
symmetry with respect to a change $A_0 \longleftrightarrow B_0$, is
related to the deviation of the Fermi function $F(E_e, Z = 1)$, caused
by the final--state electron--proton Coulomb interaction, from unity
\cite{Ivanov2012}. The parameters $X_i$ and $Y_j$ for
$i=1,2,\ldots,16$ and $j = 1,\ldots, 4$ are defined in Appendix I of
Ref.\cite{Ivanov2012}.

A dependence of our results on $\omega_m$ makes our analysis to some
extent qualitative. Nevertheless, one may show that the orders of
magnitudes of the obtained corrections are rather stable under
variations of $\omega_m$. For example, for a cut--off $\omega_m =
0.10\,{\rm MeV}$, at which the contribution of the quantum corrections
to the radiative $\beta^-$--decay of the neutron may be neglected with
respect to the classical one \cite{Jackson} proportional to
$1/\omega$, we get
\begin{eqnarray}\label{label43}
\hspace{-0.3in}\Delta X^{(s + h)}_2 &=& - 0.015290\,{\rm MeV^5},\nonumber\\
\hspace{-0.3in}\Delta X^{(s + h)}_0  &=& - 0.005509\,{\rm MeV^5},\nonumber\\
\hspace{-0.3in}\Delta X^{(s + h)}_{10} &=&  - 0.085248\,{\rm MeV^5},\nonumber\\
\hspace{-0.3in}\Delta X^{(s + h)}_{11} &=& - 0.029908\,{\rm MeV^5}.
\end{eqnarray}
These values retain the orders of magnitudes of $(\alpha/\pi)\,\Delta
X_2$ and $(\alpha/\pi)\,\Delta X_0$ and our assertion that the
radiative corrections to the lifetime of the neutron $\tau_n$ and the
proton--energy spectrum $a(T_p)$ can be described by the functions
$(\alpha/\pi)\,g_n(E_e)$ and $(\alpha/\pi)\,f_n(E_e)$. A description
of the radiative corrections to the lifetime of the neutron by the
function $(\alpha/\pi)\,g_n(E_e)$ agrees well with the results,
obtained by Gl\"uck \cite{Gluck1993}. For $\omega_m = 0.10\,{\rm MeV}$
we may introduce $X_{\rm eff} = 2.228328\,{\rm MeV^5}$. At the level
of the accuracy $10^{-5}$ this does not contradict to the results
obtained above. An approximate coincidence of the results, obtained
for $\omega_m = 0.26\,{\rm MeV}$ and $\omega_m = 0.10\,{\rm MeV}$, may
serve for a confirmation of an approximate independence of our
calculations of the proton--photon correlations of the cut--off
$\omega_m$.

\section{Comparison with results, obtained by Gl\"uck \cite{Gluck1993}}
\label{sec:gluck}

Now let us discuss our results in comparison to the results, obtained
by Gl\"uck \cite{Gluck1993}. We analyse the contributions of the
radiative corrections to the electron--antineutrino $(E_e,
\cos\theta_{e\bar{\nu}})$ distribution and the proton--energy spectrum
$a(T_p)$.

\subsection{Electron--antineutrino $(E_e,\cos\theta_{e\bar{\nu}})$ distribution}

The electron--antineutrino $(E_e,\cos\theta_{e\bar{\nu}})$
distribution of the neutron $\beta^-$--decay with unpolarised neutron
and decay electron and proton, where $\cos\theta_{e\bar{\nu}} =
\vec{k}\cdot \vec{k}_e/E k_e$ and $\vec{k}$ and $\vec{k}_e$ are the
antineutrino and electron 3--momenta, can be obtained from Eq.(6) of
Ref.\cite{Ivanov2012}. One gets
\begin{eqnarray}\label{label44}
\hspace{-0.3in}\frac{d^2 \lambda_n(E_e, \vec{k}_e, \vec{k})}{dE_e
  d\cos\theta_{e\bar{\nu}}} &=& (1 + 3
\lambda^2)\,\frac{G^2_F|V_{ud}|^2}{4\pi^3}\,(E_0 - E_e)^2
\,\sqrt{E^2_e - m^2_e}\, E_e\,F(E_e, Z = 1)\,\zeta(E_e)\,\Big(1 +
\frac{\alpha}{\pi}\,g_n(E_e)\Big)\nonumber\\
\hspace{-0.3in}&&\times\,\Big\{1 + a^{(W)}(E_e)\,\Big(1 +
\frac{\alpha}{\pi}\,f_n(E_e)\Big)\beta\,P_1(\cos\theta_{e\bar{\nu}}) -
2\,a_0\,\frac{E_e}{M}\,P_2(\cos\theta_{e\bar{\nu}}) \Big\},
\end{eqnarray}
where the function $\zeta(E_e)$, including the corrections of order
$1/M$, caused by the ``weak magnetism'' and the proton recoil, is
given in \cite{Ivanov2012}. Then, $P_1(\cos\theta_{e\bar{\nu}}) =
\cos\theta_{e\bar{\nu}}$ and $P_2(\cos\theta_{e\bar{\nu}}) = (3
\cos^2\theta_{e\bar{\nu}} - 1)/2$ are the Legendre polynomials
\cite{Jackson}. The correlation coefficient $a^{(W)}(E_e)$, containing
the leading and next--to--leading terms in the large $M$ expansion
only, is given by \cite{Ivanov2012}
\begin{eqnarray}\label{label45}
\hspace{-0.3in}a^{(W)}(E_e) &=& a_0\Big\{1 + \frac{1}{M}\,\frac{1}{(1 -
  \lambda^2)(1 + 3 \lambda^2)}\,\Big(a_1 E_0 + a_2 E_e +
a_3\frac{m^2_e}{E_e}\Big)\Big\},\nonumber\\
\hspace{-0.3in}a_1 &=& 4 \lambda (\lambda^2 + 1)(\lambda -(\kappa +
1)),\nonumber\\
\hspace{-0.3in}a_2 &=& - 26 \lambda^4 + 8(\kappa + 1)\,\lambda^3 - 20
\lambda^2 + 8(\kappa + 1)\,\lambda - 2,\nonumber\\
\hspace{-0.3in}a_3 &=& - 2 \lambda (\lambda^2 - 1)(\lambda - (\kappa +
1)).
\end{eqnarray}
We would like to accentuate that the $(E_e, \cos\theta_{e\bar{\nu}})$
distribution Eq.(\ref{label44}) is calculated in \cite{Ivanov2012} by
integrating first over the proton 3--momentum in the final state of
the continuum-state $\beta^-$--decay and the radiative
$\beta^-$--decay of the neutron, respectively. As has been pointed out
by Gl\"uck \cite{Gluck1997}, there are no proton--photon correlations
between decay protons and virtual photons in one--virtual photon
exchanges in the continuum-state $\beta^-$--decay of the neutron. The
integration over the 3--momentum of the decay proton in the radiative
$\beta^-$--decay of the neutron leads to the absence of the
proton--photon correlations to order $\alpha/\pi$ or to leading order
in the large $M$ expansion. One may show that after the integration
over the decay proton 3--momentum the proton--photon correlations in
the radiative $\beta^-$--decay of the neutron appear to order
$(\alpha/\pi)\,(E_0/M)\sim 10^{-6}$ only, which may be neglected in
comparison to contributions of order $\alpha/\pi \sim 10^{-3}$
\cite{Ivanov2012} (see also \cite{Gudkov2006}). The radiative
corrections, described by the functions $(\alpha/\pi)\,g_n(E_e)$ and
$(\alpha/\pi)\,f_n(E_e)$, are defined by the contributions of the
radiative $\beta^-$--decay of the neutron and one--virtual photon
exchanges in the continuum-state $\beta^-$--decay of the neutron (see
Eq.(D-58) of Ref.\cite{Ivanov2012})
\begin{eqnarray}\label{label46}
\hspace{-0.3in}g_n(E_e) &=&\frac{3}{2}\,{\ell n}\Big(\frac{m_p}{m_e}\Big) -
\frac{3}{8} + 2\,\Big[\frac{1}{2\beta} \,{\ell n}\Big(\frac{1 +
    \beta}{1 - \beta}\Big) - 1\Big]\Big[{\ell n}\Big(\frac{2(E_0 -
    E_e)}{m_e}\Big) - \frac{3} {2} + \frac{1}{3}\,\frac{E_0 -
  E_e}{E_e}\Big] + \frac{2}{\beta}L\Big(\frac{2\beta}{1 + \beta}\Big)\nonumber\\
\hspace{-0.3in}&+&
\frac{1}{2\beta}{\ell n}\Big(\frac{1 + \beta}{1 -
  \beta}\Big)\,\Big[(1+\beta^2) + \frac{1}{12} \frac{(E_0 -
    E_e)^2}{E^2_e} - {\ell n}\Big(\frac{1 + \beta}{1 -
    \beta}\Big)\Big] + C_{WZ}, \nonumber\\
\hspace{-0.3in}f_n(E_e) &=& \frac{2}{3}\,\frac{E_0 - E_e}{E_e}\Big(1 +
\frac{1}{8}\frac{E_0 - E_e}{E_e}\Big)\,\frac{1 - \beta^2}{\beta^2}\,
\Big[\frac{1}{2\beta}\,{\ell n}\Big(\frac{1 + \beta}{1 - \beta}\Big) -
  1\Big]- \frac{1}{12}\frac{(E_0 - E_e)^2}{E^2_e} + \frac{1 -
  \beta^2}{2\beta}\,{\ell n}\Big(\frac{1 + \beta}{1 - \beta}\Big),
\end{eqnarray}
where the constant $C_{WZ} = 10.429$ is caused by the electroweak
boson exchanges and QCD corrections (see a discussion in Appendix D of
Ref. \cite{Ivanov2012}). The contribution of $C_{WZ} = 10.249$ to the
neutron $\beta^-$--decay can be described by the parameter $\Delta_R =
(\alpha/\pi)\,C_{WZ} = 0.0238$ (see, for example,
\cite{Ivanov2012}). This value agrees well with the value $\Delta_R =
0.024$ used in \cite{Mention2011}

In Ref.\cite{Gluck1993} the radiative corrections to the $(E_e,
\cos\theta_{e\bar{\nu}})$ distribution are described by the function
$r_{e\bar{\nu}}(x, \cos\theta_{e\bar{\nu}})$, where $x = (E_e -
m_e)/(E_0 - m_e)$ and $E_0 - m_e = ((m_n - m_e)^2 - m^2_p)/2m_n =
0.7817\,{\rm MeV}$ is the $Q$--value of the neutron
$\beta^-$--decay. The function
$r_{e\bar{\nu}}(x,\cos\theta_{e\bar{\nu}})$, defined in terms of the
function $f_n(E_e)$, is
\begin{eqnarray}\label{label47}
\hspace{-0.3in}r_{e\bar{\nu}}(x,\cos\theta_{e\bar{\nu}}) =
100\,\frac{\alpha}{\pi}\,f_n(E_e).
\end{eqnarray}
Unlike the results, obtained in \cite{Gluck1993}, the function
Eq.(\ref{label47}) does not depend on $\cos\theta_{e\bar{\nu}}$. Such
an independence of $\cos\theta_{e\bar{\nu}}$ is exact to leading order
in the large $M$ expansion and caused by the integration over the
3--momentum of the decay proton that leads to the proton--photon
decorrelation.  The contribution of the radiative corrections in
\cite{Gluck1993} has been compared with the contribution, calculated
in \cite{RC1978} and given by the function
$(\alpha/\pi)\,f_n(E_e)$. As has been pointed out by Gl\"uck
\cite{RC1978}, the contribution of the radiative corrections,
calculated in \cite{Gluck1993}, is of order of magnitude larger
compared to the contribution of the radiative corrections, defined by
the function $(\alpha/\pi)\,f_n(E_e)$. In Fig.\,3 we plot the function
$r_{e\bar{\nu}}(x, \cos\theta_{e\bar{\nu}})$, given by
Eq.(\ref{label47}) and defined in terms of the function
$f_n(E_e)$. The numerical values of this function are adduced in Table
I.
\begin{figure}
\includegraphics[height=0.23\textheight]{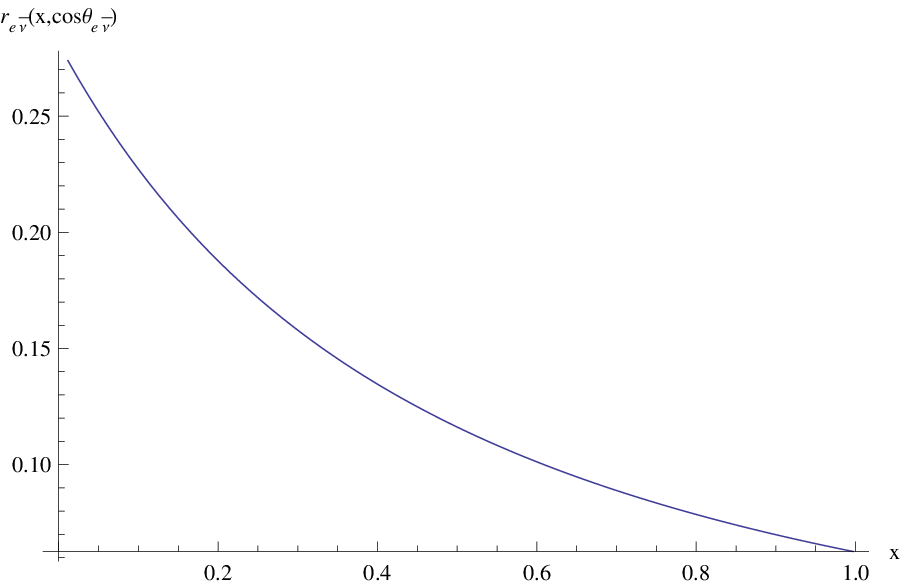}
\includegraphics[height=0.23\textheight]{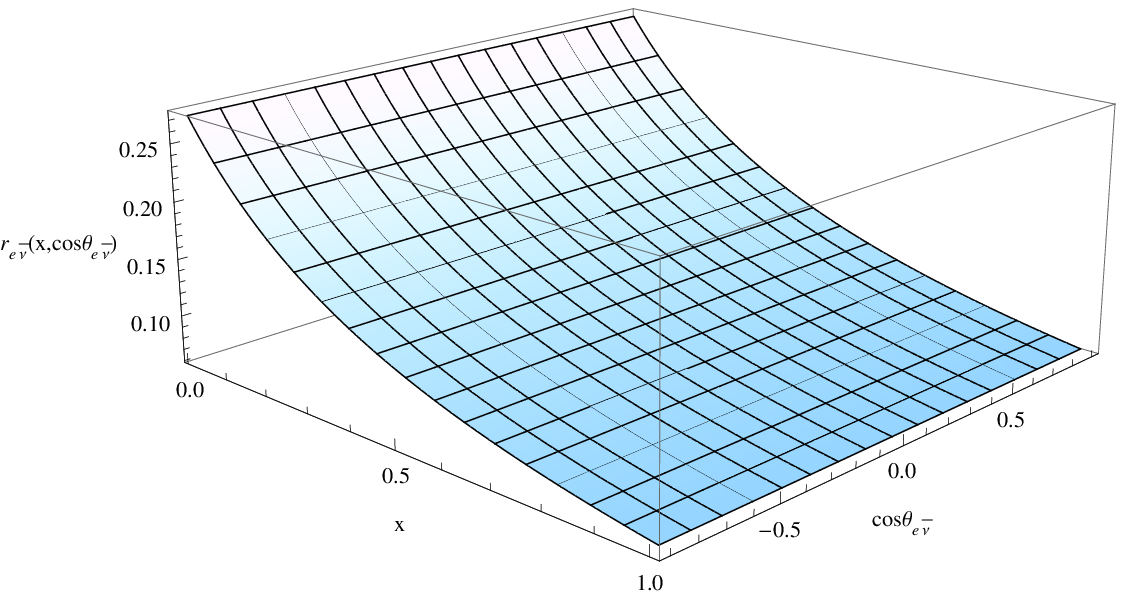}
\caption{The radiative corrections $r_{e\bar{\nu}}(x,
  \cos\theta_{e\bar{\nu}})$ to the electron--antineutrino $(E_e,
  \cos\theta_{e\bar{\nu}})$ distribution, defined in terms of the
  function $f_n(E_e)$ for $0 \le x \le 1$ with $x = (E_e - m_e)/(E_0 -
  m_e)$. }
\end{figure}
Apart from the independence of $\cos\theta_{e\bar{\nu}}$ one may see
that the values of the radiative corrections $r_{e\bar{\nu}}(x,
\cos\theta_{e\bar{\nu}})$, defined by the function $f_n(E_e)$, are
commensurable with the values of the radiative corrections, calculated
by Gl\"uck (see Table V of Ref.\cite{Gluck1993}). We would like to
emphasise that the radiative corrections, given by the function
$f_n(E_e)$, are defined by both the contributions of the soft and hard
photons \cite{Ivanov2012}.

\begin{table}[h]
\begin{tabular}{|c|c|c|c|c|c|c|c|c|c|c|c|}
\hline $x$ & $0.0$ & $0.1$ & $0.2$ & $0.3$ & $0.4$ & $0.5$ & $0.6$ &
$0.7$ & $0.8$ & $0.9$ & $1.0$\\ \hline $r_{e\bar{\nu}}(x,
\cos\theta_{e\bar{\nu}})$ & $0.28$ & $0.23$ & $0.19$ & $0.16$ & $0.14$ &
$0.12$ & $0.10$ & $0.09$ & $0.08$ & $0.070$ & $0.06$\\\hline
\end{tabular} 
\caption{The numerical values of the radiative corrections
  $r_{e\bar{\nu}}(x, \cos\theta_{e\bar{\nu}})$ to the
  electron--antineutrino $(E_e, \cos\theta_{e\bar{\nu}})$
  distribution.}
\end{table}

\subsection{Proton--energy $a(T_p)$ spectrum}

The proton--energy spectrum is defined by (see Appendix I of
Ref.\cite{Ivanov2012})
\begin{eqnarray}\label{label48}
\hspace{-0.3in}\frac{d\lambda_n(T_p)}{dT_p} = M\, (1 + 3 \lambda^2)
\frac{G^2_F|V_{ud}|^2}{4\pi^3}\,a(T_p)\,\Big(1 - b_F \Big\langle
\frac{m_e}{E_e}\Big\rangle\Big)
\end{eqnarray}
where $b_F$ is the Fierz term, $\langle m_e/E_e\rangle_{\rm SM} =
0.6556$ \cite{Ivanov2012} and $a(T_p)$ is defined by
\begin{eqnarray}\label{label49}
\hspace{-0.3in}a(T_p) = g^{(1)}_p(T_p) +
\frac{\alpha}{\pi}\,f^{(1)}_p(T_p) + a_0\Big\{g^{(2)}_p(T_p) +
\frac{\alpha}{\pi}\,\Big(f^{(2)}_p(T_p) + f^{(3)}_p(T_p)\Big)\Big\} +
b_Ff^{(4)}_p(T_p).
\end{eqnarray}
The functions $g^{(1)}_p(T_p)$ and  $g^{(2)}_p(T_p)$ are defined by  the
integrals
\begin{eqnarray}\label{label50}
\hspace{-0.3in}g^{(1)}_p(T_p) &=& \int^{(E_e)_{\rm max}}_{(E_e)_{\rm min}}
\zeta_1(E_e,T_p) \,F(E_e,
Z = 1)\,E_e\,dE_e,\nonumber\\
\hspace{-0.3in}g^{(2)}_p(T_p) &=& \int^{(E_e)_{\rm max}}_{(E_e)_{\rm
    min}} \Big(\zeta_2(E_e,T_p) + \frac{1}{1 -
  \lambda^2}\,\frac{E_0}{M}\Big)\,F(E_e, Z = 1)\,E_e\,dE_e,
\end{eqnarray}
where the functions $\zeta_1(E_e,T_p)$ and $\zeta_2(E_e,T_p)$ are
defined by Eq.(I-15) in Appendix I of Ref.\cite{Ivanov2012}. They
include the next--to--leading order corrections in the large $M$
expansion. The functions $f^{(1)}_p(T_p)$ and $f^{(2)}_p(T_p)$
determine the radiative corrections to the proton--energy
spectrum. They are given by the integrals over the electron--energy
spectrum $(E_e)_{\rm min} \le E_e \le (E_e)_{\rm max}$ in terms of the
functions $g_n(E_e)$ and $f_n(E_e)$ (see Appendix I of
Rev.\cite{Ivanov2012})
\begin{figure}
\includegraphics[height=0.23\textheight]{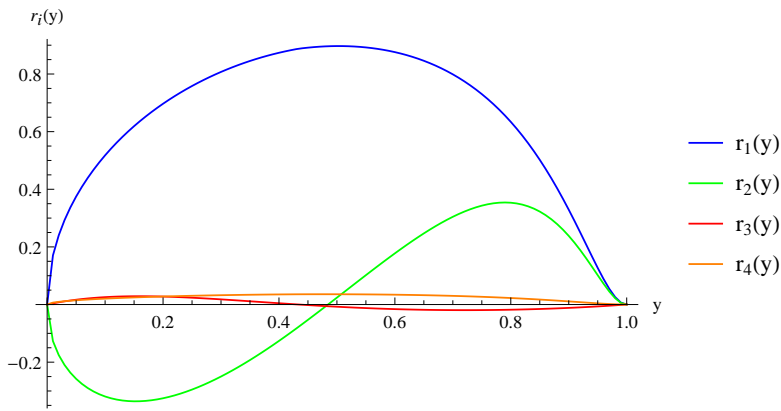}
\caption{The radiative corrections $r_1(y)$ (blue), $r_2(y)$ (green),
  $r_3(y)$ (red) and $r_4(y)$ (gold) to the proton--energy spectrum
  $a(T_p)$ for $0 \le y \le 1$ with $y = T_p/(T_p)_{\rm max}$. }
\end{figure}
\begin{eqnarray}\label{label51}
\hspace{-0.3in}f^{(1)}_p(T_p) &=& \int^{(E_e)_{\rm max}}_{(E_e)_{\rm min}}
(E_0 - E_e)\,g_n(E_e) \,F(E_e,
Z = 1)\,E_e\,dE_e,\nonumber\\
\hspace{-0.3in}f^{(2)}_p(T_p) &=& - \frac{1}{2}\int^{(E_e)_{\rm
    max}}_{(E_e)_{\rm min}} \Big((E_0 - E_e)^2 + E^2_e - m^2_e - 2
MT_p\Big)\,\Big(g_n(E_e) + f_n(E_e)\Big) \,F(E_e, Z = 1)\,dE_e.
\end{eqnarray}

\begin{table}[h]
\begin{tabular}{|c|c|c|c|c|c|c|c|c|c|c|c|}
\hline $y$ & $0.0$ & $0.1$ & $0.2$ & $0.3$ & $0.4$ & $0.5$ & $0.6$ &
$0.7$ & $0.8$ & $0.9$ & $1.0$\\ \hline $r_1(y)$ & $0.00$ & $ + 0.52$ &
$+ 0.69$ & $+ 0.81$ & $+ 0.87$ & $+ 0.90$ & $+ 0.88$ & $+ 0.80$ & $+
0.63$ & $+ 0.33$ & $0.00$\\\hline $r_2(y)$ & $0.00$ & $- 0.32$ &
$- 0.33$ & $- 0.25$ & $- 0.13$ & $+ 0.02$ & $+ 0.18$ & $+ 0.31$ & $+ 0.35$ &
$+ 0.24$ & $0.00$\\\hline $10^2 r_3(y)$ & $0.00$ & $+ 2.59$ & $+ 2.61$ &
$+ 1.80$ & $+ 0.61$ & $- 0.69$ & $- 1.76$ & $- 1.55$ & $- 1.86$ & $- 1.04$ &
$0.00$\\\hline $10^2 r_4(y)$ & $0.00$ & $+ 2.05$ & $+ 2.78$ & $+ 3.24$ &
$+ 3.51$ & $+ 3.57$ & $+ 3.42$ & $+ 3.01$ & $+ 2.26$ & $+ 1.09$ & $0.00$\\\hline
\end{tabular} 
\caption{The numerical values of the functions $r_1(y)$, $r_2(y)$ and
  $r_3(y)$, describing the radiative corrections, and the function
  $r_4(y)$, defining the contributions of the Fierz term, to the
  proton--energy spectrum $a(T_p) = a(y)$, where $T_p = y (T_p)_{\rm
    max}$. The functions $r_1(y)$, $r_2(y)$, $r_3(y)$ and $r_4(y)$ are
  obtained by the integration of the electron--proton energy $a(E_e,
  T_p)$ distribution over the energies of the decay electron
  $(E_e)_{\rm min} \le E_e \le (E_e)_{\rm max}$ and measured in ${\rm
    MeV^3}$}
\end{table}

The radiative corrections, described by the function $f^{(3)}_p(E_e)$, are
given by the proton--photon correlations in the radiative
$\beta^-$--decay of the neutron. It reads 
\begin{eqnarray*}
\hspace{-0.3in}&&f^{(3)}_p(T_p) = \Big(\Theta(T_p - T_1) - \Theta(T_p
- T_2)\Big)\int^{(E_e)_{\rm max}}_{E_0 - \omega_m}dE_e
\Bigg\{\Big\{(E_0 - E_e)^2 - \Big((E_0 - E_e)^2 - 2 M T_p + E^2_e -
m^2_e\Big)\nonumber\\
\hspace{-0.3in}&&\times\,\Big(\frac{3}{2} - \frac{1}{3}\,\frac{(E_0 -
  E_e)}{\beta^2 E_e} - \frac{1}{24}\,\frac{(E_0 - E_e)^2}{\beta^2
  E^2_e}\Big) + \frac{1}{12}\,\frac{(E_0 - E_e)^4 + 3 (2 M T_p - E^2_e +
  m^2_e)(E_0 - E_e)^2}{\beta^2
  E^2_e}\nonumber\\
\hspace{-0.3in}&& + \frac{(2 M T_p - E^2_e + m^2_e)(E_0 -
  E_e)}{\beta^2 E_e} \Big\}\,\Big[\frac{1}{\beta}\,{\ell
    n}\Big(\frac{1 + \beta}{1 - \beta}\Big) - 2\Big] -
\frac{1}{3}\,\frac{(E_0 - E_e)^3}{E_e}\,\frac{1}{\beta}\,{\ell
  n}\Big(\frac{1 + \beta}{1 - \beta}\Big)- E_e (E_0 -
E_e)
\end{eqnarray*}
\begin{eqnarray}\label{label52}
\hspace{-0.3in}&&\times\,\Big[\frac{3 - \beta^2}{\beta}\,{\ell
    n}\Big(\frac{1 + \beta}{1 - \beta}\Big) - 6\Big] \Bigg\}\, F(E_e,
Z = 1)\, dE_e + \Big(\Theta(T_p - T_2) - \Theta(T_p - (T_p)_{\rm
  max})\Big)\int^{(E_e)_{\rm max}}_{(E_e)_{\rm min}}dE_e
\Bigg\{\Big\{(E_0 - E_e)^2\nonumber\\
\hspace{-0.3in}&& - \Big((E_0 - E_e)^2 - 2 M T_p + E^2_e -
m^2_e\Big)\,\Big(\frac{3}{2} - \frac{1}{3}\,\frac{(E_0 - E_e)}{\beta^2
  E_e} - \frac{1}{24}\,\frac{(E_0 - E_e)^2}{\beta^2 E^2_e}\Big)+
\frac{(2M T_p - E^2_e + m^2_e)(E_0 - E_e)}{\beta^2 E_e} \nonumber\\
\hspace{-0.3in}&& + \frac{1}{12}\,\frac{(E_0 - E_e)^4 + 3 (2 M T_p -
  E^2_e + m^2_e)(E_0 - E_e)^2}{\beta^2
  E^2_e}\Big\}\,\Big[\frac{1}{\beta}\,{\ell n}\Big(\frac{1 + \beta}{1
    - \beta}\Big) - 2\Big] - \frac{1}{3}\,\frac{(E_0 -
  E_e)^3}{E_e}\,\frac{1}{\beta}\,{\ell n}\Big(\frac{1 + \beta}{1 -
  \beta}\Big)\nonumber\\
\hspace{-0.3in}&& - E_e (E_0 - E_e)\,\Big[\frac{3 -
    \beta^2}{\beta}\,{\ell n}\Big(\frac{1 + \beta}{1 - \beta}\Big) -
  6\Big] \Bigg\}\, F(E_e, Z = 1) + \Big(\Theta(T_p) - \Theta(T_p -
T_1)\Big)\int^{(E_e)_{\rm max}}_{(E_e)_·{\rm
    min}}dE_e\nonumber\\
\hspace{-0.3in}&&\times\,\Bigg\{\frac{(E_0 - E_e)^2 - 2 M T_p + E^2_e
  - m^2_e}{E_e}\,g^{(2)}_{\beta^-_c\gamma}(E_e,\omega_m) + \Delta f_a
(E_e, \sqrt{2 M T_p})\, \sqrt{2 M T_p(E^2_e - m^2_e)}\Bigg\}\,E_e
\,F(E_e, Z = 1)\nonumber\\
\hspace{-0.3in} && + \Big(\Theta(T_p - T_1) - \Theta(T_p -
T_2)\Big)\int^{E_0 - \omega_m}_{(E_e)_·{\rm min}}dE_e\,\Bigg\{\frac{(E_0 - E_e)^2
  - 2 M T_p + E^2_e -
  m^2_e}{E_e}\,g^{(2)}_{\beta^-_c\gamma}(E_e,\omega_m) + \Delta
f_a(E_e, \sqrt{2M T_p})\nonumber\\
\hspace{-0.3in}&&\times\,\sqrt{2 M T_p(E^2_e - m^2_e)}\Bigg\}\,E_e
\,F(E_e, Z = 1),
\end{eqnarray}
where $T_j = k^2_j/2M$ and $j = 1,2$ with $k_1$ and $k_2$ defined in
Eq.(\ref{label21}). Then, the function $f^{(4)}_p(T_p)$ is given by
\begin{eqnarray}\label{label53}
\hspace{-0.3in}f^{(4)}_p(T_p) = m_e\int^{(E_e)_{\rm max}}_{(E_e)_{\rm
    min}} (E_0 - E_e)\,F(E_e, Z = 1)\,dE_e
\end{eqnarray}
Following Gl\"uck \cite{Gluck1993} we introduce the functions
\begin{eqnarray}\label{label54}
\hspace{-0.3in}r_1(y) &=&
100\,\frac{\alpha}{\pi}\,f^{(1)}_p(T_p)\;,\;r_2(y) =
100\,\frac{\alpha}{\pi}\,f^{(2)}_p(T_p),\nonumber\\
\nonumber\\
\hspace{-0.3in}r_3(y) &=& 100\,\frac{\alpha}{\pi}\,f^{(3)}_p(T_p)
\;,\; r_4(y) = 100\,\frac{\alpha}{\pi}\,f^{(4)}_p(T_p),
\end{eqnarray}
where $y = (E_p - m_p)/((E_p)_{\rm max} - m_p) = T_p/(T_p)_{\rm
  max}$ \cite{Gluck1993} with $(T_p)_{\rm max} = (E^2_0 -
m^2_e)/2M$. The functions $r_1(y)$, $r_2(y)$, $r_3(y)$ and $r_4(y)$
are plotted in Fig.\,4. The numerical values of these functions for $0
\le y \le 1$ are adduced in Table II. The contribution of the function
$r_4(y)$ to $a(T_p)$ is proportional to $0.01\,b_F\,\pi/\alpha = 4.31\,
b_F$.

One may see that the contributions of the radiative corrections,
described by the function $r_3(y)$ and induced by the proton--photon
correlations, are of order of magnitude smaller compared with the
contributions of the radiative corrections, described by the function
$r_2(y)$ and determined by the functions $g_n(E_e)$ and
$f_n(E_e)$. Since in addition the function $r_3(y)$ changes a sign
around $y \simeq 0.45$ or $T_p \simeq 0.340\,{\rm keV}$, the result of
the integration of $r_3(y)$ over the proton energy spectrum $0 \le k_p
\le (k_p)_{\rm max}$ or $0 \le y \le 1$ confirms our assertion,
concerning a negligibility of the term, proportional to $a_0$, in the
proton recoil angular distribution Eq.(\ref{label38}).

\begin{figure}
\includegraphics[height=0.23\textheight]{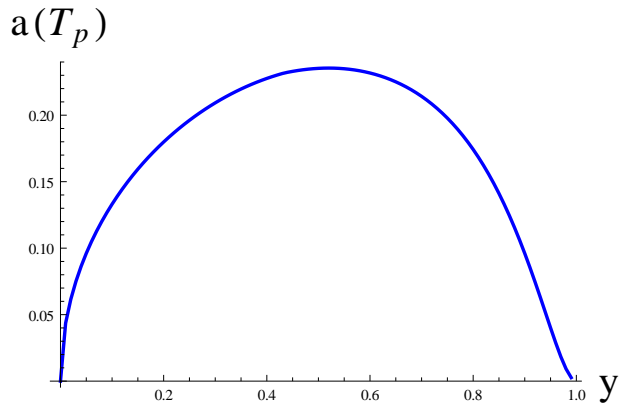}
\caption{The proton--energy spectrum $a(T_p)$. The maximal value
  $a(T_p)_{\rm max} = 0.235\,{\rm MeV^3}$ is located around $T_p =
  0.391\,{\rm keV}$ or $y = 0.52$. }
\end{figure}

In Fig.\,5 we plot the proton--energy spectrum $a(T_p)$. A maximum of
the proton--energy spectrum $a(T_p)$ is located around $T_p \simeq
0.391\,{\rm keV}$ or around $y \simeq 0.52$ (see also
\cite{Stratowa1978}). One may see that for the decay protons with
energies $0.40 \le y \le 0.55$ or $0.300\,{\rm keV} \le T_p \le
0.413\,{\rm keV}$ the contributions of the radiative corrections
$r_3(y)$, caused by the proton--photon correlations can be fairly
neglected in comparison to the contributions of the radiative
corrections $r_2(y)$, defined by the functions $g_n(E_e)$ and
$f_n(E_e)$.

Of course, for the decay protons, detected from the energy region
$0.413\,{\rm keV} \le T_p \le 0.680\,{\rm keV}$ or $0.55 \le y \le
0.90$ \cite{Byrne1990,Byrne2002} with the aim to measure the
contributions of the Fierz term \cite{Nab2009,Nab2012,Konrad2011},
defined by scalar and tensor interactions beyond the SM
\cite{Ivanov2012}, one should take into account the contributions of
the proton--photon correlations, given by the function $r_3(y)$. A
recent estimate of the Fierz term $b_F = 3.2(2.3)\times 10^{-3}$ has
been carried out in \cite{Ivanov2009}. The Fierz term, multiplied by
$4.31\,r_4(y)$, gives the contributions of order $10^{-4}$ to the
proton--energy spectrum $a(T_p)$.

Comparing the function $r_2(y)$ with the function $r_p(y)$ (see Table
IV of Ref.\cite{Gluck1993}), one may see that these functions differ
in a sign and a behaviour at $y \to 1$. Indeed, the function $r_2(y)$
is negative in the interval $0.00 \le y \le 0.45$ and changes sign at
$y \simeq 0.45$. whereas the function $r_p(y)$ is positive in the
interval $0.00 \le y \le 0.62$ and changes sign at $y \simeq
0.62$. Then, the function $r_2(y)$ is positive in the interval $0.45
\le y \le 1$ and vanishes at $y = 1$, whereas the function $r_p(y)$ is
negative in the interval $0.62 \le y \le 1$ and does not vanish at $y
= 1$. 

The vanishing of the function $r_2(y)$ is obvious, since this function
is defined by the integral over the electron energies $(E_e)_{\rm min}
\le E_e \le (E_e)_{\rm max}$ and $(E_e)_{\rm min} = (E_e)_{\rm max} =
(E^2_0 + m^2_e)/2E_0$ for $y = 0$ and $(E_e)_{\rm min} = (E_e)_{\rm
  max} = E_0$ for $y = 1$ or for $T_p = 0$ and $T_p = (T_p)_{\rm
  max}$, respectively. The sign of the function $r_2(y)$ is defined by
the sign of the function $f(E_e, T_p) = - ((E_0 - E_e)^2 + E^2_e -
m^2_e - 2 MT_p)/2$ in the integrand of the integral over the electron
energies, which is changed at $T_p = (E^2_0 - 2 m^2_e)/4M$ and $E_e =
E_0/2$.

\subsection{Electron--proton energy $a(E_e, T_p)$ distribution}

We have analysed in detail the radiative corrections to the
proton--energy spectrum $a(T_p)$. Since the proton--energy spectrum
$a(T_p)$ is related to the electron--proton energy distribution
$a(E_e, T_p)$ as \cite{Ivanov2012}
\begin{eqnarray}\label{label55}
\hspace{-0.3in}a(T_p) = \int^{(E_e)_{\rm max}}_{(E_e)_{\rm min}}a(E_e,
T_p)\,F(E_e, Z = 1)\,E_e \, dE_e,
\end{eqnarray}
a dominance of the radiative corrections, described by the functions
$g_n(E_e)$ and $f_n(E_e)$, with respect to the radiative corrections,
caused by the proton--photon correlations, is also valid for the
electron--proton energy distribution.

\section{Acknowledgements}

We are grateful to Hartmut Abele and Manfried Faber for fruitful
discussions and Ferenz Gl\"uck for calling our attention to his paper
\cite{Gluck1997} and the problem of the proton recoil energy and
angular distribution of the neutron $\beta^-$--decay, caused the
proton--photon correlations in the radiative $\beta^-$--decay of the
neutron.  This work was supported by the Austrian ``Fonds zur
F\"orderung der Wissenschaftlichen Forschung'' (FWF) under the
contracts I689-N16, I534-N20 PERC and I862-N20 and by the Russian
Foundation for Basic Research under the contract No. 11-02-91000
-ANF$_-$a.


\begin{thebibliography}{9}
\bibitem{Ivanov2012} 
A. N. Ivanov, M. Pitschmann, and
  N. I. Troitskaya, arXiv: 1212.0332 [hep--ph].
\bibitem{Sirlin1967}
A. Sirlin,
Phys. Rev. {\bf 164}, 1767 (1967).
\bibitem{Abers1969}
E. S. Abers, D. A. Dicus, R. E. Norton, and H. R. Queen,
Phys. Rev. {\bf 167}, 1461 (1968).
\bibitem{Shann1971}
R. T. Shann, 
 Cimento A {\bf 5}, 591 (1971).
\bibitem{Gluck1997}
F. Gl\"uck,
Computer Physics Communications {\bf 101}, 223 (1997).
\bibitem{Christian1978}
R. Christian and H. K\"uhnelt,
Acta Phys. Austriaca {\bf 49}, 229 (1978).
\bibitem{Treiman1958}
S. B. Treiman,
Phys. Rev. {\bf 110}, 448 (1958).
\bibitem{Abele2008} 
H. Abele, Progr. Part. Nucl. Phys. {\bf 60}, 1
  (2008).
\bibitem{Nico2009}
J. S. Nico, 
J. Phys. G: Nucl. Part. Phys. {\bf 36}, 104001 (2009).
\bibitem{Gluck1993}
F. Gl\"uck,
Phys. Rev. C {\bf 47}, 2840 (1993).
\bibitem{HMF72} 
{\it Handbook of Mathematical Functions with Formulas,
  Graphs, and Mathematical Tables}, ed. by M. Abramowitz and
  I. A. Stegun, Tenth Printing with corrections, National Bureau of
  Standards Applied Mathematics Series $\bullet$ 55, p.1004, 1972.
\bibitem{PolyLog1}
  K. Mitchell, Phil. Mag.  {\bf 40}, 351 (1949).
\bibitem{PolyLog2}
 E. S. Ginsberg and
  D. Zaborowski, Communications of the ACM, {\bf 18}, 200 (1975).
\bibitem{PolyLog3}
L. Lewin, in {\it Polylogarithms and associated functions},
North Holland, New York, 1981.
\bibitem{Gluck1996}
F. Gl\"uck,
Phys. Lett. B {\bf 376}, 25 (1996).
\bibitem{Wolfram}
S. Wolfram,
in {\it Mathematica 8.0},  Mathematica Version 3, Addison--Wesley, Bonn 1997.
\bibitem{Jackson}
J. D. Jackson,
in {\it Classical Electrodynamics}, John Wiley $\&$ Sons, Inc.,
New York, Chapter 15, 1962.
\bibitem{Gudkov2006} 
V. Gudkov, G. I. Greene, and J. R. Calarco,
  Phys. Rev. C {\bf 73}, 035501 (2006).
\bibitem{Mention2011}
G. Mention {\it et al.},
Phys. Rev. D {\bf 83}, 073006 (2011) and references therein.
\bibitem{RC1978}
A. Garc$\acute{\rm i}$a and M. Maya,
Phys. Rev. D {\bf 17}, 1376 (1978).
\bibitem{Stratowa1978}
Chr. Stratowa, R. Dobrozemsky, and P. Weinzierl,
Phys. Rev. D {\bf 18}, 3970 (1978).
\bibitem{Byrne1990}
J. Byrne {\it et al.},
Phys. Rev. Lett. {\bf 65}, 289 (1990).
\bibitem{Byrne2002}
J. Byrne {\it et al.}, 
J. Phys. G: Nucl. Part. Phys. {\bf 28}, 1325 (2002).
\bibitem{Nab2009}
D. Po${\check{\rm c}}$ani${\acute{\rm c}}$ {\it et al.},
Nucl. Instr. Meth. in Phys. Res. A {\bf 611}, 211 (2009).
\bibitem{Nab2012}
S. Bei\ss ler {\it et al.},
{\it Neutron beta decay studies with Nab}, arXiv: 1209.4663 [nucl-ex].
\bibitem{Konrad2011} 
G. Konrad, PhD Thesis, {\it Measurements of the
  proton recoil spectrum in neutron beta decay with the spectrometer
  {\rm aSPECT}: study of systematic effects}, Mainz, August 2011.
\bibitem{Ivanov2009}
M. Faber {\it et al.},
Phys. Rev. C {\bf 80}, 035503 (2009).
\end{thebibliography}
\end{document}